\documentclass[longauth]{aa}  
\usepackage{wasysym}
\usepackage{hyperref}
\hypersetup{ breaklinks=true,colorlinks=true,urlcolor=blue, linkcolor=blue,  citecolor=blue, bookmarksopen=true}
\usepackage{graphicx}
\usepackage{txfonts}
\usepackage{tikz}
\usetikzlibrary{shapes,arrows}
\graphicspath{{modelling/}{./}}
\newcommand{\Lsun}{\mathrm{L}_\odot}
\newcommand{\Msun}{\mathrm{M}_\odot}

\begin{document} 

\titlerunning{The asymmetric inner disk of HD 163296 in the eyes of MATISSE}
   \title{The asymmetric inner disk of the Herbig Ae star HD 163296 in the eyes of VLTI/MATISSE: evidence for a vortex?\thanks{Based on observations made with ESO Telescopes at the La Silla Paranal Observatory under program IDs 0103.D-0294 and 0103.D-0153.}}

   \author{
J.~Varga\inst{\ref{inst_L},\ref{inst_K}}\and
M.~Hogerheijde\inst{\ref{inst_L},\ref{inst_P}}\and
R.~van~Boekel\inst{\ref{inst_H}}\and
L.~Klarmann\inst{\ref{inst_H}}\and
R.~Petrov\inst{\ref{inst_O}}\and
L.B.F.M.~Waters\inst{\ref{inst_Ra},\ref{inst_U}}\and
S.~Lagarde\inst{\ref{inst_O}}\and
E.~Pantin\inst{\ref{inst_Pa}}\and
Ph.~Berio\inst{\ref{inst_O}}\and
G.~Weigelt\inst{\ref{inst_B}}\and
S.~Robbe-Dubois\inst{\ref{inst_O}}\and
B.~Lopez\inst{\ref{inst_O}}\and
F.~Millour\inst{\ref{inst_O}}\and
J.-C.~Augereau\inst{\ref{inst_I}}\and
H.~Meheut\inst{\ref{inst_O}}\and
A.~Meilland\inst{\ref{inst_O}}\and
Th.~Henning\inst{\ref{inst_H}}\and
W.~Jaffe\inst{\ref{inst_L}}\and
F.~Bettonvil\inst{\ref{inst_A}}\and
P.~Bristow\inst{\ref{inst_Garch}}\and
K.-H.~Hofmann\inst{\ref{inst_B}}\and
A.~Matter\inst{\ref{inst_O}}\and
G.~Zins\inst{\ref{inst_E}}\and
S.~Wolf\inst{\ref{inst_R}}\and
F.~Allouche\inst{\ref{inst_O}}\and
F.~Donnan\inst{\ref{inst_L}}\and
D.~Schertl\inst{\ref{inst_B}}\and
C.~Dominik\inst{\ref{inst_P}}\and
M.~Heininger\inst{\ref{inst_B}}\and
M.~Lehmitz\inst{\ref{inst_H}}\and
P.~Cruzal\`ebes\inst{\ref{inst_O}}\and
A.~Glindemann\inst{\ref{inst_Garch}}\and
K.~Meisenheimer\inst{\ref{inst_H}}\and
C.~Paladini\inst{\ref{inst_E}}\and
M.~Sch\"oller\inst{\ref{inst_Garch}}\and
J.~Woillez\inst{\ref{inst_Garch}}\and
L.~Venema\inst{\ref{inst_A}}\and
E.~Kokoulina\inst{\ref{inst_O}}\and
G.~Yoffe\inst{\ref{inst_H}}\and
P.~\'Abrah\'am\inst{\ref{inst_K},\ref{inst_Eo}}\and
S.~Abadie\inst{\ref{inst_E}}\and
R.~Abuter\inst{\ref{inst_Garch}}\and
M.~Accardo\inst{\ref{inst_Garch}}\and
T.~Adler\inst{\ref{inst_H}}\and
T.~Ag\'ocs\inst{\ref{inst_A}}\and
P.~Antonelli\inst{\ref{inst_O}}\and
A.~B\"ohm\inst{\ref{inst_H}}\and
C.~Bailet\inst{\ref{inst_O}}\and
G.~Bazin\inst{\ref{inst_Garch}}\and
U.~Beckmann\inst{\ref{inst_B}}\and
J.~Beltran\inst{\ref{inst_E}}\and
W.~Boland\inst{\ref{inst_L}}\and
P.~Bourget\inst{\ref{inst_E}}\and
R.~Brast\inst{\ref{inst_Garch}}\and
Y.~Bresson\inst{\ref{inst_O}}\and
L.~Burtscher\inst{\ref{inst_L}}\and
R.~Castillo\inst{\ref{inst_E}}\and
A.~Chelli\inst{\ref{inst_O}}\and
C.~Cid\inst{\ref{inst_E}}\and
J.-M.~Clausse\inst{\ref{inst_O}}\and
C.~Connot\inst{\ref{inst_B}}\and
R.D.~Conzelmann\inst{\ref{inst_Garch}}\and
W.-C.~Danchi\inst{\ref{inst_O},\ref{inst_Go}}\and
M.~De Haan\inst{\ref{inst_A}}\and
M.~Delbo\inst{\ref{inst_O}}\and
M.~Ebert\inst{\ref{inst_H}}\and
E.~Elswijk\inst{\ref{inst_A}}\and
Y.~Fantei\inst{\ref{inst_O}}\and
R.~Frahm\inst{\ref{inst_Garch}}\and
V.~G\'amez Rosas\inst{\ref{inst_L}}\and
A.~Gabasch\inst{\ref{inst_Garch}}\and
A.~Gallenne\inst{\ref{inst_Co},\ref{inst_Wa},\ref{inst_Sa}}\and
E.~Garces\inst{\ref{inst_E}}\and
P.~Girard\inst{\ref{inst_O}}\and
F.Y.J.~Gont\'e\inst{\ref{inst_Garch}}\and
J.C.~Gonz\'alez Herrera\inst{\ref{inst_Garch}}\and
U.~Graser\inst{\ref{inst_H}}\and
P.~Guajardo\inst{\ref{inst_E}}\and
F.~Guitton\inst{\ref{inst_O}}\and
X.~Haubois\inst{\ref{inst_E}}\and
J.~Hron\inst{\ref{inst_V}}\and
N.~Hubin\inst{\ref{inst_Garch}}\and
R.~Huerta\inst{\ref{inst_E}}\and
J. W.~Isbell\inst{\ref{inst_H}}\and
D.~Ives\inst{\ref{inst_Garch}}\and
G.~Jakob\inst{\ref{inst_Garch}}\and
A.~Jask\'o\inst{\ref{inst_A},\ref{inst_K}}\and
L.~Jochum\inst{\ref{inst_E}}\and
R.~Klein\inst{\ref{inst_H}}\and
J.~Kragt\inst{\ref{inst_A}}\and
G.~Kroes\inst{\ref{inst_A}}\and
S.~Kuindersma\inst{\ref{inst_A}}\and
L.~Labadie\inst{\ref{inst_C}}\and
W.~Laun\inst{\ref{inst_H}}\and
R.~Le Poole\inst{\ref{inst_L}}\and
C.~Leinert\inst{\ref{inst_H}}\and
J.-L.~Lizon\inst{\ref{inst_Garch}}\and
M.~Lopez\inst{\ref{inst_E}}\and
A.~M\'erand\inst{\ref{inst_Garch}}\and
A.~Marcotto\inst{\ref{inst_O}}\and
N.~Mauclert\inst{\ref{inst_O}}\and
T.~Maurer\inst{\ref{inst_H}}\and
L.H.~Mehrgan\inst{\ref{inst_Garch}}\and
J.~Meisner\inst{\ref{inst_L}}\and
K.~Meixner\inst{\ref{inst_H}}\and
M.~Mellein\inst{\ref{inst_H}}\and
L.~Mohr\inst{\ref{inst_H}}\and
S.~Morel\inst{\ref{inst_O}}\and
L.~Mosoni\inst{\ref{inst_Zs},\ref{inst_K}}\and
R.~Navarro\inst{\ref{inst_A}}\and
U.~Neumann\inst{\ref{inst_H}}\and
E.~Nu{\ss}baum\inst{\ref{inst_B}}\and
L.~Pallanca\inst{\ref{inst_E}}\and
L.~Pasquini\inst{\ref{inst_Garch}}\and
I.~Percheron\inst{\ref{inst_Garch}}\and
J.-U.~Pott\inst{\ref{inst_H}}\and
E.~Pozna\inst{\ref{inst_Garch}}\and
A.~Ridinger\inst{\ref{inst_H}}\and
F.~Rigal\inst{\ref{inst_A}}\and
M.~Riquelme\inst{\ref{inst_E}}\and
Th.~Rivinius\inst{\ref{inst_E}}\and
R.~Roelfsema\inst{\ref{inst_A}}\and
R.-R.~Rohloff\inst{\ref{inst_H}}\and
S.~Rousseau\inst{\ref{inst_O}}\and
N.~Schuhler\inst{\ref{inst_E}}\and
M.~Schuil\inst{\ref{inst_A}}\and
A.~Soulain\inst{\ref{inst_Sy}}\and
P.~Stee\inst{\ref{inst_O}}\and
C.~Stephan\inst{\ref{inst_E}}\and
R.~ter Horst\inst{\ref{inst_A}}\and
N.~Tromp\inst{\ref{inst_A}}\and
F.~Vakili\inst{\ref{inst_O}}\and
A.~van Duin\inst{\ref{inst_A}}\and
J.~Vinther\inst{\ref{inst_Garch}}\and
M.~Wittkowski\inst{\ref{inst_Garch}}\and
F.~Wrhel\inst{\ref{inst_H}}
}

   \institute{
Leiden Observatory, Leiden University, Niels Bohrweg 2, NL-2333 CA Leiden, the Netherlands\label{inst_L}\\\email{varga@strw.leidenuniv.nl} \and
Konkoly Observatory, Research Centre for Astronomy and Earth Sciences,  Konkoly Thege Mikl\'os \'ut 15-17, H-1121 Budapest, Hungary\label{inst_K} \and
Anton Pannekoek Institute for Astronomy, University of Amsterdam, Science Park 904, 1090 GE Amsterdam, The Netherlands\label{inst_P} \and
Max Planck Institute for Astronomy, K\"onigstuhl 17, D-69117 Heidelberg, Germany\label{inst_H} \and
Laboratoire Lagrange, Universit\'e C\^ote d'Azur, Observatoire de la C\^ote d'Azur, CNRS, Boulevard de l'Observatoire, CS 34229, 06304 Nice Cedex 4, France\label{inst_O} \and
Institute for Mathematics, Astrophysics and Particle Physics, Radboud University, P.O. Box 9010, MC 62 NL-6500 GL Nijmegen, the Netherlands\label{inst_Ra} \and
SRON Netherlands Institute for Space Research, ﻿Sorbonnelaan 2, NL-3584 CA Utrecht, the Netherlands\label{inst_U} \and
AIM, CEA, CNRS, Universit\'e Paris-Saclay, Universit\'e Paris Diderot, Sorbonne Paris Cit\'e, F-91191 Gif-sur-Yvette, France\label{inst_Pa} \and
Max-Planck-Institut f\"ur Radioastronomie, Auf dem H\"ugel 69, D-53121 Bonn, Germany\label{inst_B} \and
Univ. Grenoble Alpes, CNRS, IPAG, 38000, Grenoble, France\label{inst_I} \and
NOVA Optical IR Instrumentation Group at ASTRON (Netherlands)\label{inst_A} \and
European Southern Observatory, Karl-Schwarzschild-Stra\ss e 2, 85748 Garching, Germany\label{inst_Garch} \and
European Southern Observatory, Alonso de Cordova 3107, Vitacura, Santiago, Chile\label{inst_E} \and
Institut f\"ur Theoretische Physik und Astrophysik, Christian-Albrechts-Universit\"at zu Kiel, Leibnizstra{\ss}e 15, 24118, Kiel, Germany\label{inst_R} \and
ELTE E\"otv\"os Lor\'and University, Institute of Physics, P\'azm\'any P\'eter s\'et\'any 1/A, 1117 Budapest, Hungary\label{inst_Eo} \and
NASA Goddard Space Flight Center, Astrophysics Division, Greenbelt, MD 20771, USA\label{inst_Go} \and
Departamento de Astronom\'ia, Universidad de Concepci\'on, Casilla 160-C, Concepci\'on, Chile\label{inst_Co} \and
Nicolaus Copernicus Astronomical Centre, Polish Academy of Sciences, Bartycka 18, 00-716 Warszawa, Poland\label{inst_Wa} \and
Unidad Mixta Internacional Franco-Chilena de Astronom\'ia (CNRS UMI 3386), Departamento de Astronom\'ia, Universidad de Chile, Camino El Observatorio 1515, Las Condes, Santiago, Chile\label{inst_Sa} \and
Department of Astrophysics, University of Vienna, T\"urkenschanzstrasse 17, A-1180 Vienna, Austria\label{inst_V} \and
I. Physikalisches Institut, Universit\"at zu K\"oln, Z\"ulpicher Str. 77, 50937, K\"oln, Germany\label{inst_C} \and
Zselic Park of Stars, 064/2 hrsz., 7477 Zselickisfalud, Hungary\label{inst_Zs} \and
Sydney Institute for Astronomy, School of Physics, A28, The University of Sydney, NSW 2006, Australia\label{inst_Sy} 
             }

   \date{Received September 15, 1996; accepted March 16, 1997}

  \abstract
   {The inner few au region of planet-forming disks is a complex environment. High angular resolution observations have a key role in understanding the disk structure and the dynamical processes at work.}
   {In this study we aim to characterize the mid-infrared brightness distribution of the inner disk of the young intermediate-mass star HD 163296, from early VLTI/MATISSE observations, taken in the L- and N-bands. We put special emphasis on the detection of potential disk asymmetries.}
   {We use simple geometric models to fit the interferometric visibilities and closure phases. Our models include a smoothed ring, a flat disk with inner cavity, and a 2D Gaussian. The models can account for disk inclination and for azimuthal asymmetries as well. We also perform numerical hydro-dynamical simulations of the inner edge of the disk.}
   {Our modeling reveals a significant brightness asymmetry in the L-band disk emission. The brightness maximum of the asymmetry is located at the NW part of the disk image, nearly at the position angle of the semimajor axis. The surface brightness ratio in the azimuthal variation is $3.5 \pm 0.2$. Comparing our result on the location of the asymmetry with other interferometric measurements, we confirm that the morphology of the $r<0.3$~au disk region is time-variable. We propose that this asymmetric structure, located in or near the inner rim of the dusty disk, orbits the star. For the physical origin of the asymmetry, we tested a hypothesis where a vortex is created by Rossby wave instability, and we find that a unique large scale vortex may be compatible with our data. 
   The half-light radius of the L-band emitting region is $0.33\pm 0.01$~au, the inclination is ${52\degr}^{+5\degr}_{-7\degr}$, and the position angle is $143\degr \pm 3\degr$. Our models predict that a non-negligible fraction of the L-band disk emission originates inside the dust sublimation radius for $\mu$m-sized grains. Refractory grains or large ($\gtrsim 10\ \mu$m-sized) grains could be the origin for this emission. N-band observations may also support that the innermost disk ($r\lesssim 0.6$~au) lacks small silicate grains, in agreement with our findings from L-band data.
   }
   {}

               
   \maketitle
%

\section{Introduction}
\label{sec:intro}
In the first few million years of stellar evolution, stars are surrounded by a gas- and dust-rich circumstellar disk. These protoplanetary or planet-forming disks are dynamic and complex environments, with many processes at work: turbulence, gas accretion to the central star, outflows, disk winds, dust grain growth and settling, myriad of chemical reactions, etc. Planet-forming disks are also the cradles of planets. In the recent years our knowledge on the structure of planet-forming disks significantly increased, mostly thanks to high angular resolution facilities like ALMA, VLT/SPHERE, and VLTI. It has been discovered that planet-forming disks have substructures: rings, gaps, spiral arms, and asymmetric features are commonly found in them \citep{vanBoekel2017,Andrews2018,Avenhaus2018,Huang2018}. These substructures are on tens of au scale. However, disk structure in the inner few au region, where terrestrial planets form, is much less known. Infrared (IR) interferometric facilities like VLTI are complementary to (sub-)mm instruments, as they can provide valuable constraints on $\lesssim 1$~au spatial scales, although their capability to reveal fine structure is somewhat limited, due to the sparse baseline coverage. Near-infrared interferometry reveals the inner rim of the dusty disk ($r \sim 0.1 - 1$~au), while in the mid-infrared we can see a larger disk region extending to a few au. Inner holes and gaps are the most common substructures that are observed by IR interferometry, especially in (pre-)transitional disks \citep[e.g.,][]{Menu2014,Matter2016}. Asymmetric disk features are also detected in a number of cases \citep{Kraus2009_RCrA,Kraus2013,Weigelt2011,Panic2014,Jamialahmadi2018,Kluska2020}, even with time-variable morphology \citep{Kluska2016}. A notable result of statistical studies based on near-IR interferometric observations is the establishment of a correlation between the stellar luminosity and the radius of the disk inner rim \citep{Monnier2005,Eisner2007,Renard2010,Dullemond2010,Lazareff2017,Perraut2019}. This relation arises from the fact that there is a dust-free zone near the star where the temperature is above the dust sublimation temperature ($\sim$$1500$~K). The size of this region is determined by the luminosity of the star. The size-luminosity relation has been also confirmed by mid-IR observations, although the mid-IR disk emitting region shows a greater structural variety than the near-IR one \citep{vanBoekel2005_flaring,Monnier2009,Menu2015,Millan-Gabet2016,Varga2018}.  Also in active galactic nuclei, the near-IR size luminosity relation is much stricter than the mid-IR one, showing this is really a universal behavior \citep{Burtscher2013}. 

In this study, we focus on HD 163296, a well-studied $7-10$~Myr old \citep{Vioque2018,Setterholm2018} Herbig Ae star (A1Vep spectral type) in Sagittarius at a distance of $101.2 \pm 1.2$~pc \citep{Bailer-Jones2018}. A recent estimation of its stellar parameters gave a stellar luminosity of $16\ \Lsun$ and a stellar mass of $1.9\ \Msun$ \citep{Setterholm2018}. Its disk has been spatially resolved at many wavelengths from the near-IR to the mm. At mm-wavelengths, the ALMA image shows an inclined disk with numerous sharp rings and annular gaps \citep{Huang2018}. These features are located between $10$~au and $155$~au from the star. There are two asymmetric features as well: one is crescent-like asymmetry near the inner edge of the $r = 67$~au ring towards SE, the other is located inside the $r = 10$~au gap towards SW. IR interferometric instruments resolved the inner few au region of the disk. The half-light radius of the N-band emitting region is found to be around $1.2$~au \citep{Menu2015,Varga2018}. In the near-IR, the bulk of emission comes from inside $r \approx 0.3$~au, partly from inside the dust sublimation radius \citep[e.g.,][]{Benisty2010_HD163296,Setterholm2018,Perraut2019}. \citet{Millan-Gabet2016} fitted Keck near- and mid-IR interferometric data with a 2-rim disk model, and found $0.39$~au for the inner, and $1.1$~au for the outer rim radius (rescaled using the current distance estimate). Several authors reported an asymmetric brightness distribution of the inner disk \citep{Lazareff2017,Kluska2020}. \citet{Lazareff2017} found that an azimuthally modulated ring with a radius of $0.25$~au gives a good fit to PIONIER H-band data. In a recent study \citet{Kluska2020} performed a direct image reconstruction from PIONIER data. The resulting image is centrally peaked, without an inner cavity. The nondetection of the cavity may be due to the limitations in resolution ($\sim$$0.2$~au). \citet{Setterholm2018} fitted an extensive set of H- and K-band interferometric data, and found that a Gaussian gives a better fit than a narrow ring. Recent GRAVITY observations clearly indicate a variable morphology: new reconstructed images from two observation campaigns (one in 2018, the other in 2019) show an asymmetric arc-like feature which changed its position between the two epochs (GRAVITY Collaboration, in prep.). The nature of this variable feature is not yet clear. HD 163296 has a jet which ejects variable amounts of material on time scales of years, causing near-IR photometric variability \citep[e.g.,][]{Ellerbroek2014}. \citet{Sitko2008} discussed the variability of the inner rim and the connection to the outflow/jet. They conclude that the variations in the $1-5\ \mu$m flux indicate structural changes in the disk region near the dust sublimation zone.

As the example of HD 163296 shows, disk substructure and asymmetries in the terrestrial planet forming zone are still not well characterized. To improve this, more detailed observations with mas-scale resolution are needed. Recently, MATISSE, the new interferometric instrument on the VLTI opened up two new wavelength ranges, L- and M-bands ($3-5\ \mu$m), for interferometry \citep[][in prep.]{Lopez2014_MATISSE,Matter2016_MATISSE,Lopez2021}. MATISSE simultaneously observes the N band, like its predecessor, MIDI. In all three bands, it provides 6 simultaneous baselines and 3 independent closure phases. Due to this wide wavelength coverage, MATISSE is both sensitive to the dust sublimation zone and to the disk regions in the terrestrial planet forming zone.

In this paper, we present the first MATISSE L- and N-band observations of HD 163296, with the aim to model the disk with simple geometric models, and characterize its asymmetry. The structure of the paper is as follows. In Sect.~\ref{sec:obs} we describe the observations and in Sect.~\ref{sec:dataproc} the data reduction. In Sect.~\ref{sec:model} we explain our models to interpret the data. In Sect.~\ref{sec:res} we present our results. In Sect.~\ref{sec:discussion} we compare our results to the literature, and try to find the physical origin of the disk asymmetry. Finally, in Sect.~\ref{sec:summary} we summarize our findings.

 	\begin{table*}
		\caption{Overview of VLTI/MATISSE observations of HD 163296. $\tau_0 $ is the atmospheric coherence time. LDD is the estimated angular diameter of the calibrator. $\vartheta_\mathrm{max}$ is the resolution corresponding to the longest baseline in the array. Bands indicate the band(s) that the calibrator was used for.}
		\begin{center}
			\label{tab:obs}
			\begin{tabular}{c c c c c c c c c c c}
				\hline
				\hline
				\multicolumn{5}{c}{Target} & & & \multicolumn{4}{c}{Calibrator} \\
				\hline
				Date and time (UTC)& Seeing & $\tau_0$ & Stations & Array & \multicolumn{2}{c}{$\vartheta_\mathrm{max}$ (mas)} & Name & LDD & Bands & Time (UTC) \\
				 & ($''$) & (ms) &  & & L & N & & (mas)&  \\
				\hline
				2019-03-23 08:41	& 0.45 & 9.7 &	A0-B2-D0-C1	& small & 11 & 33 & $\delta$ Sgr & 5.86 & LN &  08:53\\ 
				2019-05-06 08:19 & 0.7 & 4.6 & K0-G1-D0-J3 & large & 3 & 8 & HD 156637 & 2.16 & L &  08:07\\
				 	&  &  &	&  & & & $\delta$ Sgr & 5.86 & N & 08:38 \\ 
				2019-06-26 06:26 & 2.5 & 0.9 & A0-B2-D0-C1 & small & 11 & 34 & HD 165135 & 3.46 & LN & 07:25 \\
				2019-06-29 07:07 & 0.9 & 2.4 & A0-B2-D0-C1 & small & 11 & 35 & HD 165135 & 3.46 & LN & 07:38 \\
				\hline
			\end{tabular}
		\end{center}
	\end{table*}

  \begin{figure}
   \centering
   \includegraphics[width=\hsize]{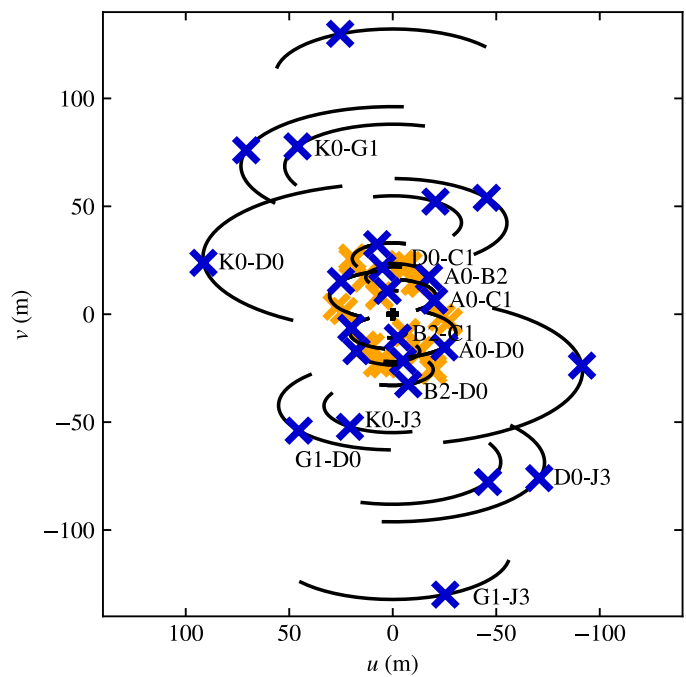}
      \caption{The $uv$-coverage of our observations. Blue crosses represent the data from 2019 March and May, orange crosses represent the data from 2019 June.  }
         \label{fig:uv}
   \end{figure}

\section{Observations}
\label{sec:obs}

MATISSE is the latest four-telescope interferometer on the Very Large Telescope Interferometer (VLTI) at the European Southern Observatory (ESO) Paranal Observatory \citep[][in prep.]{Lopez2014_MATISSE, Matter2016b, Matter2016_MATISSE,Lopez2021}. The instrument operates in three wavelength domains: the L-band ($2.9-4.2\ \mu$m), the M-band ($4.6-5\ \mu$m) and the N-band ($8-13\ \mu$m). It measures visibility, differential phase, closure phase, correlated flux, and total flux. MATISSE has two detectors, one for the L- and M-bands, the other for the N-band. A typical observation consists of 2 sky exposures, followed by an exposure cycle of 4 non-chopped interferometric exposures, each taking 1 min. During the interferometric exposures, L-, M-, and N-band interferometric data, along with L- and M-band total flux data are taken. The non-chopped exposure cycle can be repeated, e.g., to reach a better signal-to-noise ratio. The 4 exposures within a cycle are not identical: there are two beam commuting devices (BCDs) at the entrance of the instrument which commute the beams coming from the telescopes. A single non-chopped exposure corresponds to one of the 4 different BCD configurations. Beam commutation helps to reduce instrumental effects, especially for the phase signal \citep{Millour2008}. N-band total flux is optionally recorded after the interferometric exposures. During these, so-called photometric, observations 8 chopped exposures are taken. The photometric exposures contain L- and M-band interferometric data, along with L-, M-, and N-band total flux data. A MATISSE exposure consists of several hundred (in L- and M-bands) or several thousand (in N-band) frames. A frame is a single integration, with a detector integration time on the order of $0.1$~s in L-, M-bands, and $20$~ms in N-band. Typical sensitivity limits in low spectral resolution mode on the Auxiliary Telescopes (ATs) are are $1-1.5$~Jy in L-band, $4-6$~Jy and in N-band. For more details on the instrument performance we refer to \citet[][in prep.]{Lopez2021}.
   
Data on HD 163296 were taken with the ATs between March and June 2019 as part of the MATISSE guaranteed time observing campaign. M-band data were not recorded. We used low spectral resolution both in L-band ($\lambda/\delta\lambda \approx 34$), and in N-band ($\lambda/\delta\lambda \approx 30$). During all observations, except in March, we recorded two non-chopped exposure cycles. The observations in March and June were obtained using the short AT baselines, while in May using a large AT configuration. While the atmospheric conditions during the March and May observations were good to excellent, the data in June were obtained in unfavorable weather, with an atmospheric coherence time $\tau_0 \lesssim 2.5$~ms. Table~\ref{tab:obs} provides an overview of the observations obtained with MATISSE. The data sets cover the baselines between $10$ and $130$~m, corresponding to $2.9-37.1\,\mathrm{M}\lambda$ spatial frequency range in the L-band. Fig.~\ref{fig:uv} shows the $uv$-coverage of our observations. The corresponding spatial scales are $2.8-36$~mas ($0.3-3.6$~au) in the L-band, and $8.5-110$~mas ($0.9-11$~au) in the N-band\footnote{We compute the spatial scale (resolution) as $\vartheta = \lambda / \left(2B_\mathrm{p}\right)$, where $\lambda$ is the wavelength, and $B_\mathrm{p}$ is the projected baseline length. This is the usual convention in optical-IR interferometry.}. During March and May no N-band photometric observations were taken. 
  
Calibrator stars were observed right after the science observations. For the observations with the small AT array, a single calibrator was selected for both bands. However, for the medium and large AT configurations, the stellar diameter of the calibrator  should be less than $\sim$$3$~mas in case of the L-band, and $\sim$$9$~mas in case of the N-band. These criteria ensure that calibration errors caused by the uncertainty in the calibrator diameter remain small. As most $\lesssim$$3$~mas diameter stars are too faint to be suitable N-band calibrators, we chose distinct calibrators for the two MATISSE bands. Further important selection criteria are that the angular separation and airmass differences between the target and calibrator should be as small as possible\footnote{Thus, we required that the azimuth difference between target and calibrator should be $<30^\circ$, and the airmass difference should be $<0.2$.}. We used both the Mid-infrared stellar Diameters and Fluxes compilation Catalogue (MDFC, \citealp{Cruzalebes2019}), and the older calibrator catalog for VLTI/MIDI \citep{vanBoekelPhD2004} for selecting the \mbox{calibrators}. 
   
\section{Data processing}
\label{sec:dataproc}

     \begin{figure}
   \centering
   \includegraphics[width=\hsize]{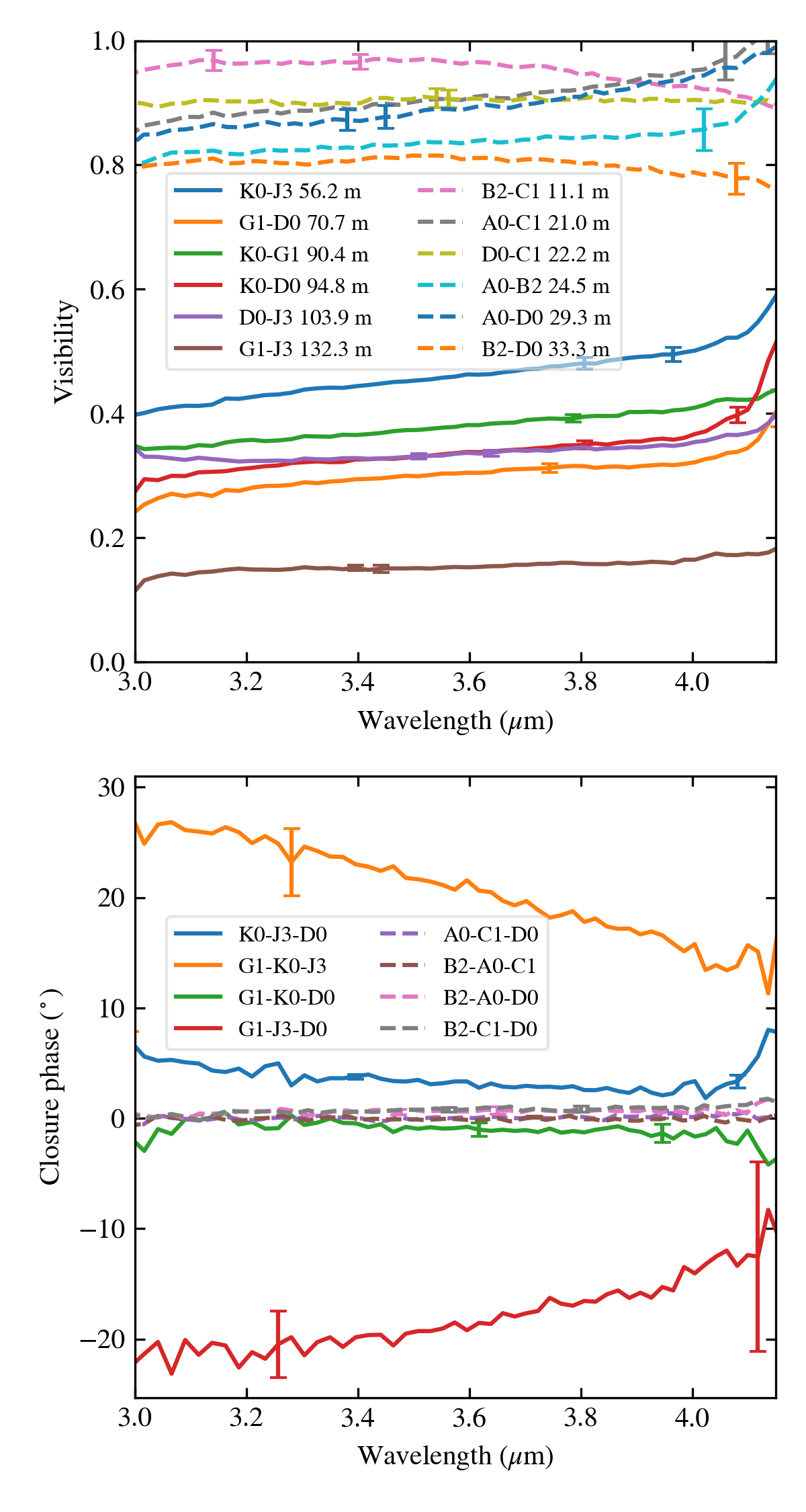}
      \caption{Final L-band calibrated data products from the 2019 March (dashed lines) and May (solid lines) observations: spectrally resolved absolute visibilities (top), and closure phases (bottom). Error bars are indicated at a few random locations.
             }
         \label{fig:caldataL}
   \end{figure}
           \begin{figure}
   \centering
   \includegraphics[width=\hsize]{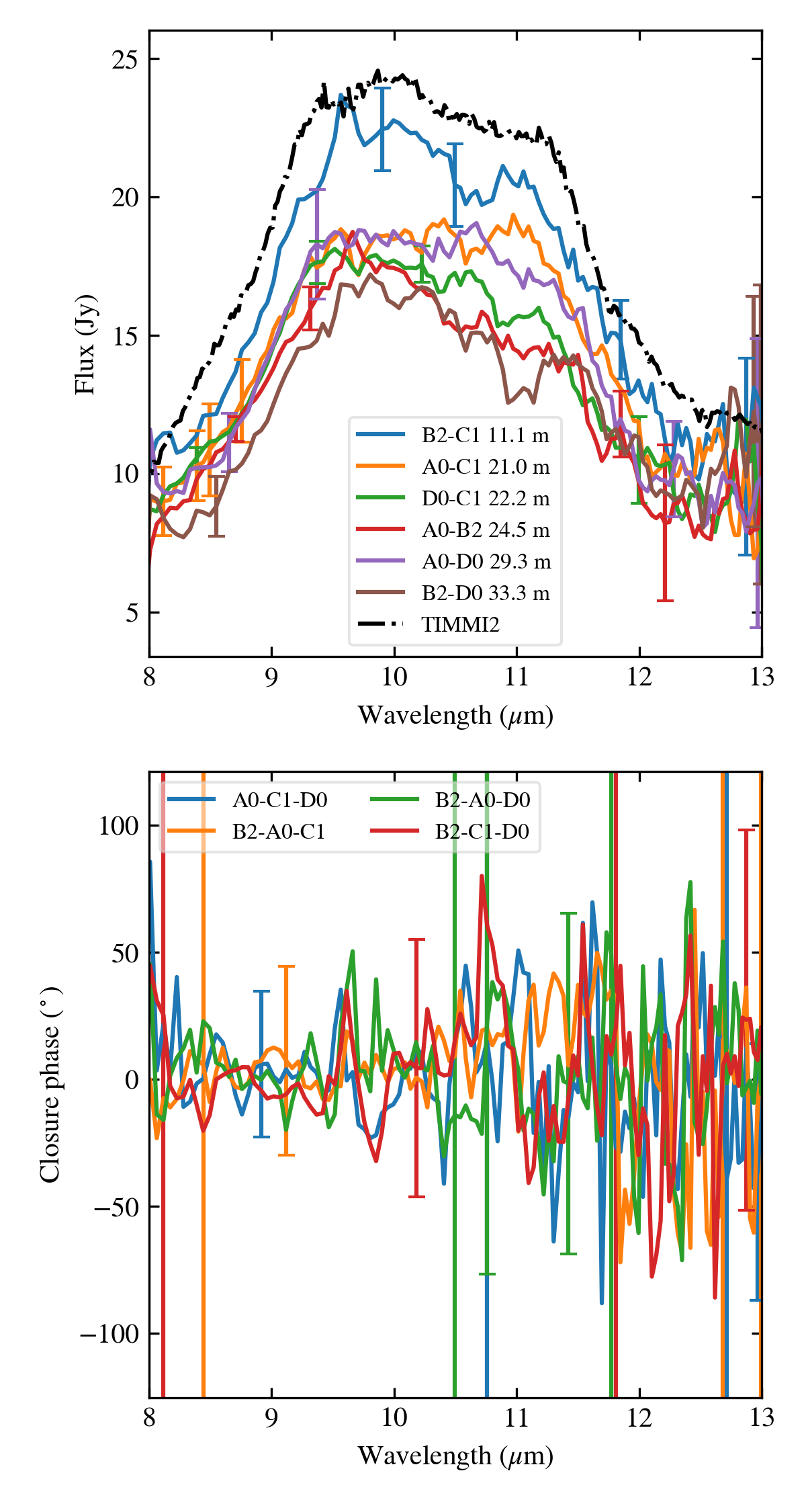}
      \caption{Final N-band calibrated data products from the 2019 March observation (solid lines): correlated spectra (top), and closure phases (bottom). Error bars are indicated at a few random locations. For comparison, we plot the total spectrum on the top panel (dash-dot black line), measured with the TIMMI2 instrument \citep{vanBoekel2005}. 
             }
         \label{fig:caldataN}
   \end{figure}
   
\subsection{Data reduction and calibration}
Our data processing consists of the following stages: data reduction, calibration, averaging, and error analysis. In Fig.~\ref{fig:flowchart} in the appendix we present a flow chart of the general workflow. We reduced the data with the standard MATISSE data reduction pipeline DRS version 1.5.0 \citep{Millour2016}.  The pipeline takes the Fourier-transform of the interferograms frame-by-frame, extracts the complex correlated flux for each baseline, and then averages the correlated flux over all frames. The pipeline provides two alternative methods for this averaging, one is the coherent mode, the other is the incoherent mode. The difference is that in coherent mode the average of the correlated fluxes is taken linearly, while in incoherent mode the squared correlated fluxes are averaged. The coherent method needs an estimation for the optical path delay, which is used to correct the phase of the complex correlated flux prior averaging. Visibilities are calculated by dividing the averaged correlated flux by the average total flux. For more information we refer to \citet{Millour2016}\footnote{The pipeline recipes are explained in detail in the MATISSE pipeline user manual, available at \url{ftp://ftp.eso.org/pub/dfs/pipelines/instruments/matisse/matisse-pipeline-manual-1.5.1.pdf}}. For the L-band data we applied the incoherent method to calculate visibilities. In N-band, however, we utilized the coherent method to obtain correlated fluxes\footnote{The non-default data reduction options were compensate = "pb,rb,nl,if,bp,od" in L-band; and corrFlux=TRUE, useOpdMod=TRUE, spectralBinning=7 in N-band.}. There are several reasons for doing this: 1) L-band correlated flux and total flux data are recorded simultaneously. Thus, by using visibility the influence of  variable flux levels caused by atmospheric variations is largely eliminated. However, in N-band the photometric exposures are recorded separately from the interferometric ones. This introduces an additional error on the visibility, because of the unknown change in the transfer function between the interferometric and photometric exposures. The amount of this uncertainty largely depends on the stability of the atmosphere. 2) The coherent estimator, applied to N-band data, has a significant gain in signal-to-noise ratio (SNR), compared to the incoherent method.  In L-band the coherent and incoherent estimators provide comparable performance. 3) The N-band total flux of our target ($F_{\mathrm{tot},\,N} \approx 20$~Jy) is below the sensitivity limit of MATISSE with the ATs ($F_{\mathrm{tot},\,N} = 25-30$~Jy), thus N-band visibilities cannot be estimated.
   
The next step in our data processing is the calibration. For the L-band data we applied the usual visibility calibration as implemented in the DRS. This method can be expressed in the following way:
   \begin{equation}
       V(\lambda) = V_\mathrm{raw}(\lambda) / T(\lambda),
   \end{equation}
where $V(\lambda)$ is the calibrated visibility of the science target, $V_\mathrm{raw}(\lambda)$ is the raw visibility of the science target, and $T(\lambda)$ is the transfer function, derived from the calibrator observation. The transfer function is raw visibility of the calibrator after correcting for its spatial extent. For the N-band data we performed direct flux calibration of the raw correlated fluxes. As this method is not part of the DRS, we developed our own tools to calibrate these data. The direct flux calibration is described by the following equations:
   \begin{eqnarray}
       F_{\mathrm{corr,}\nu}(\lambda) = F^\mathrm{raw}_{\mathrm{corr,}\nu}(\lambda)/T_{\mathrm{corr,}\nu}(\lambda),\\
       T_{\mathrm{corr,}\nu}(\lambda) = \frac{F^\mathrm{cal,raw}_{\mathrm{corr,}\nu}(\lambda)}{ F^\mathrm{cal}_{\mathrm{tot,}\nu}(\lambda) V^\mathrm{cal}(\lambda)}.
   \end{eqnarray}
Here $F_{\mathrm{corr,}\nu}(\lambda)$ is the calibrated correlated flux of the science target, $F^\mathrm{raw}_{\mathrm{corr,}\nu}(\lambda)$ is the raw correlated flux of the science target, and $T_{\mathrm{corr,}\nu}(\lambda)$ is the transfer function for flux\footnote{The flux density values are expressed in unit frequency ($\nu$), in units of Jy.}. $T_{\mathrm{corr,}\nu}(\lambda)$ is estimated by dividing the raw correlated flux of the calibrator, $F^\mathrm{cal,raw}_{\mathrm{corr,}\nu}(\lambda)$, by the modeled correlated flux of the calibrator. The latter quantity is calculated as the product of the total spectrum ($F^\mathrm{cal}_{\mathrm{tot,}\nu}(\lambda)$) and the visibility ($V^\mathrm{cal}(\lambda)$) of the calibrator. Both $F^\mathrm{cal}_{\mathrm{tot,}\nu}(\lambda)$ and $V^\mathrm{cal}(\lambda)$ are usually inferred from fitting stellar atmosphere models to photometric data. We obtained the N-band model spectra of the calibrators ($\delta$ Sgr, HD 165135) from the MIDI calibrator database \citep{vanBoekelPhD2004}. $V^\mathrm{cal}(\lambda)$ is calculated assuming a uniform disk geometry where the angular diameter of the star is taken from the MDFC. The L- and N-band closure phases were calibrated using the DRS, in the frame of the visibility calibration recipe. The calibration results in several calibrated data-sets, one for each exposure. To get the final calibrated data we take the average of these data-sets.
   
Proper characterization of the uncertainties in the processed data is highly important for correct interpretation. In the Appendix~\ref{sec:error} we perform a detailed quality assessment of the calibrated data. Using the results of this analysis, we exclude the N-band data from May, and all June data from the modeling. This choice is also supported by the evaluation of the L-band transfer function for each of the observing nights, shown in Figures~\ref{fig:TF_L_03_22}-\ref{fig:TF_L_06_28}. Additionally, we set conservative lower limits on the total uncertainties, which are $0.03$ for the L-band visibility, $1^\circ$ for the L-band closure phase, and $8\%$ for the N-band correlated flux\footnote{We did not determine a global uncertainty limit for the N-band closure phase, because we do not use these data in the modeling.}. These values will be used in our modeling.
   
\subsection{Final calibrated data}

Our final calibrated data sets from 2019 March and May, shown in Figs.~\ref{fig:caldataL} and \ref{fig:caldataN}, consist of the L-band absolute visibilities, the N-band correlated spectra, and the L- and N-band closure phases\footnote{Our data products are available in OIFITS format at the Optical interferometry DataBase (OiDB) at the Jean-Marie Mariotti Center (\url{http://oidb.jmmc.fr}).}. All these data are spectrally resolved. L-band visibilites, ranging from $0.15$ (at 132~m baseline) to $0.96$ (at 11~m baseline), indicate that the disk is well resolved. The closure phases with the small AT array ($B_\mathrm{p} < 33$~m) are mostly within $\pm1^\circ$, but on the longer baselines we can see a large signal. These MATISSE closure phases (in the $\pm 30^\circ$ range) are significantly larger than the H-band PIONIER \citep[$\pm 15^\circ$, ][]{Kluska2020}, and K-band GRAVITY values \citep[$-10^\circ ... +4^\circ$, ][]{Perraut2019}\footnote{Older K-band observations by VLTI/AMBER show closure phases in the range of $-60^\circ \dots +40^\circ$, although with very large ($\sim 30^\circ$) uncertainties \citep{Setterholm2018}.}, but smaller than CHARA K-band closure phases which were found up to 90 degrees \citep[measured at much higher spatial frequencies than the MATISSE data,][]{Setterholm2018}. This closure phase signal indicates that there is a significant departure from centro-symmetry in the L-band brightness distribution at spatial scales $< 6.4$~mas ($< 0.7$~au) corresponding to the longer baselines ($B_\mathrm{p} > 56$~m). No spectral features (e.g., the $3.3\ \mu$m polycyclic aromatic hydrocarbon (PAH) band, arising from C-H stretch resonance) can be seen in the L-band data. PAH emission strongly depends on the ultraviolet radiation field. Unlike the radiative equilibrium attained by (sub-)micron sized dust grains, the absorption and re-emission of radiation by PAHs and very small grains (VSGs) does not correspond to an equilibrium temperature \citep[e.g.,][]{Visser2007}. The absence of PAH emission in the spectrum of HD 163296 suggests that the dust grains emitting in the L-band are most likely not PAHs or VSGs, but larger grains in thermal equilibrium. However, the presence of small dehydrogenated grains still cannot be excluded.
 
The N-band correlated spectra are in the flux range of $10-20$~Jy, showing a prominent silicate spectral emission feature. For comparison, we plot a total spectrum measured with the TIMMI2 instrument \citep{vanBoekel2005}, on Fig.~\ref{fig:caldataN}. The shape of the silicate feature is similar to earlier observations with MIDI \citep{Varga2018} and Spitzer \citep{Juhasz2012}, indicating the presence of large amorphous and various-sized crystalline silicate grains.   The shape of most MIDI correlated spectra are similar, suggesting that the crystallinity does not vary much with radius. However, the silicate feature is absent from the MIDI correlated spectra on baselines $>74$~m (corresponding to a resolution of $\vartheta<1.5$~au). With the current MATISSE data we cannot probe these spatial scales, because the N-band data taken with the large AT-array are unreliable (as described in the previous section). Looking at only the small array MATISSE data, the correlated flux decreases with baseline length, indicating that the object is mildly resolved. We do not see any significant deviation from zero in the N-band closure phases, as the signal is quite noisy. We note that the N-band performance of MATISSE is still being assessed, and developments in the data reduction pipeline are expected to increase the N-band data quality. 
   
\section{Interferometric modeling}
\label{sec:model}
\begin{table*}
\caption{List of the parameters in our models.
}
\begin{center}
	\label{tab:model}
\begin{tabular}{l p{4cm} p{4.5cm} p{4.5cm}}
\hline \hline
Parameter & Ring model & Flat disk model & 2D Gaussian model \\
 & L-band & L-band & L- \& N-band \\
\hline
$\theta$ & \multicolumn{3}{c}{position angle of the major axis (East from North)}\\
$\cos\,i$  & \multicolumn{3}{c}{axis ratio}\\
$HWHM_\mathrm{Gaussian}$ & - & - & radius (HWHM) of Gaussian \\
$F_\mathrm{tot}$ & - & - & total flux \\
$F_\mathrm{tot, \star}$ & - & - & total flux of the central star \\
$f_\star$  & \multicolumn{2}{c}{flux ratio of the central star} & - \\
$R_\mathrm{in}$ & ring radius & inner radius & - \\
$FWHM_\mathrm{kernel}$ & kernel width (FWHM) & - & - \\
$A_\mathrm{mod}$  & \multicolumn{2}{c}{amplitude of the azimuthal modulation} & - \\
$\phi_\mathrm{mod}$ & \multicolumn{2}{c}{phase angle of the azimuthal modulation (from the major axis)} & -\\
$q$  & - & power-law exponent of the temperature gradient & - \\
$\log\,f$  & \multicolumn{3}{c}{logarithm of the error underestimation fraction}\\
\hline
\end{tabular}
\end{center}
\end{table*}
	
There are two main approaches for the interpretation of IR interferometric data: one is direct image reconstruction, the other is model fitting, either with geometric or radiative transfer models. The first method requires dense $uv$-sampling, and since we only have $uv$-points from two telescopes configurations, we opt for modeling. Our main goal is to characterize the L- and N-band brightness distributions with simple geometric models,  by means of deriving some key parameters such as half-light radius, disk orientation, size of the inner cavity, and location of the asymmetry. Simple models are capable to describe the disk emission in a small wavelength range, thus we model the L-band and N-band data separately. Additionally, we select different models for the different bands, so that each model is best suited either for the L-band or for the N-band data-set. We use the following models in this work:
\begin{itemize}
    \item An asymmetric ring model, based on \citet{Lazareff2017} (L-band).
    \item An asymmetric flat disk model with inner cavity (L-band). 
    \item A 2D Gaussian model (L- and N-band).
\end{itemize}

All models account for the disk inclination and position angle. The ring model and the flat disk model have the same number of free parameters. Stellar flux is also taken into account: the central star is a point at the origin with a fixed flux ratio ($f_\star$) with respect to the total flux. The model parameters are listed in Table~\ref{tab:model} with short explanations.

\subsection{Asymmetric ring model}

As mentioned in Sect.~\ref{sec:intro}, \citet{Lazareff2017} modeled the H-band data of HD~163296 with an asymmetric ring model. In the near-IR, ring models usually work well for Herbig Ae/Be disks, as the bulk of emission comes from the brightly illuminated inner rim of the dust disk. The L-band disk emission could still be dominated by the inner rim. Therefore, the model we adopt in this work, that is based on \citet{Lazareff2017}, is an elliptical ring with a first-order azimuthal modulation. The ring is assumed centered on the star, and its apparent ellipticity represents its inclination on the sky plane. An azimuthal modulation is introduced to be able to interpret the nonzero closure phases. The model image is convolved with an elliptical pseudo-Lorentzian kernel \citep{Lazareff2017} to account for the radial thickness of the ring. This model can also represent a centrally peaked brightness distribution, if the kernel size is significantly larger than the ring diameter. The fitted parameters are listed in Table~\ref{tab:model}. A minor difference between our model than that of \citet{Lazareff2017} is the prescription of the azimuthal modulation:
\begin{equation}
\label{eq:mod}
    F_\mathrm{mod}\left(r,\phi\right) = F_0\left(r,\phi\right) \left(1 + A_\mathrm{mod} \cos\left(\phi - \phi_\mathrm{mod}\right) \right),
\end{equation}
where $r$ is the radius, $\phi$ is the polar angle, $A_\mathrm{mod}$ is the amplitude of the modulation, $\phi_\mathrm{mod}$ is the modulation angle (with respect to the major axis of the ellipse), $F_0$ is the unmodulated image, and $F_\mathrm{mod}$ is the modulated image. This is equivalent to the first order azimuthal modulation ($m=1$) in \citet{Lazareff2017}. A further difference from \citet{Lazareff2017} is that we do not include a spatially extended halo component in our model, as the data do not suggest the presence of such structure (visibilites at short baselines are very close to $1$)\footnote{The fact that no extended component is needed in our model may suggest that the physical origin of the halo emission seen at shorter wavelengths in many Herbig stars is the scattering of the stellar light.} For more details on the model geometry we refer to \citet{Lazareff2017}.

\subsection{Flat disk model}

As an alternative  model to describe the L-band brightness distribution, we apply a physically motivated flat disk geometry, based on \citet{Menu2015} and \citet{Varga2018}, which has an inner cavity, a sharp inner rim at $R_\mathrm{in}$ radius, and a gradually decreasing brightness distribution outside the rim. 
 The brightness distribution is determined by the temperature structure, which has a power-law radial profile. The disk surface emits black-body radiation~: 
\begin{equation}
\label{eq:I_nu}
    I_\nu \left( r \right) \propto B_\nu  \left( T_\mathrm{in} \left( \frac{r}{R_\mathrm{in}} \right)^{-q} \right),
\end{equation}
where $T_\mathrm{in}$ the temperature at the inner radius, and $q$ is the power-law exponent \citep[e.g.,][]{Hillenbrand1992}. Following \citet{Dullemond2001}, we treat the inner edge of the disk as an optically thick wall, and set $T_\mathrm{in}$ to the local blackbody equilibrium temperature. In this case $T_\mathrm{in}$ is only dependent on the luminosity of the central star ($L_\star$)~:
\begin{equation}
T_\mathrm{in} = \left( \frac{L_\star}{4\pi\sigma R_\mathrm{in}^2} \right)^{1/4},
\end{equation}
where $\sigma$ is the Stefan–Boltzmann constant. For a discussion on the validity of this formula, we refer to \citet{Dullemond2010}. We are fitting visibilities, which constrain the shape of the surface brightness distribution but not its absolute level. Therefore, we do not constrain the proportionality factor in Eq.~\ref{eq:I_nu}. We note that this factor is roughly equivalent to the optical depth of the warm disk surface layer, from which most of the emission we see arises. The model has two parameters describing the structure of the disk, $R_\mathrm{in}$ and $q$. The power-law exponent $q$ accounts for the width of the emitting region. We apply azimuthal modulation in the same way as for the ring model, defined in Eq.~\ref{eq:mod}.

\subsection{2D Gaussian model}

As the object is weakly resolved on the short AT baselines in the N-band, and the corresponding closure phases are consistent with zero signal within the error bars,  the previously introduced models cannot be well constrained by the N-band data. Here we require a simple symmetric model, with fewer number of parameters. Thus, we choose a centrally symmetric 2D elongated Gaussian to describe the N-band disk emission. We fit this model to the N-band correlated fluxes. The main fitted parameter is the half width half maximum of the Gaussian ($HWHM_\mathrm{Gaussian}$). The total flux of the system ($F_\mathrm{tot}$) is also a fitted parameter. We have three fixed parameters in this model: the flux of the central star ($F_\mathrm{tot, \star}$), the axis ratio ($\cos i$), and the position angle of the major axis of the Gaussian ($\theta$), regarded as the disk position angle. The latter two values were taken from ALMA observations \citep{Huang2018}.

We also apply the 2D Gaussian model to the L-band data, with the inclusion of an additional offset point source accounting for asymmetry. This model is described in more detail in the Appendix~\ref{sec:gauss_model}.

\subsection{Fitting procedure}
\label{sec:modelfit}
In order to find the best-fit parameters, we perform $\chi^2$-minimization. In L-band we calculate the $\chi^2$ for the visibilities and for the closure phases separately, and then we sum these two values to get the total $\chi^2$ value. We do not apply any weighting either to the visibility or to the closure phase when calculating the total $\chi^2$. By using a simple sum we did not experience any quality difference between the fits to visibilities and the fits to closure phases. In N-band we calculate the $\chi^2$ only for the correlated flux.

Our modeling procedure consists of the following steps:
\begin{itemize}
    \item We generate the model image.
        \begin{itemize}
        \item First, we create a Cartesian coordinate grid.
        \item Then we rotate the grid by $\theta$, and scale the coordinates in the $x$ direction by the axis ratio $\cos i$.
        \item We generate the brightness distribution on the rotated and scaled grid. Due to the coordinate transform, the isophotes will be ellipses.
        \item In case of the ring and flat disk models, we introduce azimuthal modulation, according to Eq.~\ref{eq:mod}.
        \item In case of the ring model, we convolve the image with an ellipsoidal pseudo-Lorentzian kernel, with a full width half maximum (FWHM) of the major axis of $\mathrm{FWHM}_\mathrm{kernel}$. The orientation and axis ratio of the kernel is same as of the ring.
        \end{itemize}
    \item We then take the discrete Fourier-transform of the image at the $uv$-coordinates of the data. 
    \item We calculate the model visibilities (or correlated fluxes), and closure phases.
    \item Finally, we compare the model with the data by estimating the $\chi^2$.
\end{itemize}

The optimization of the model parameters is accomplished with a Python implementation of Goodman \& Weare’s Markov chain Monte Carlo (MCMC) ensemble sampler, called {\em emcee} \citep{Foreman-Mackey2013a,Foreman-Mackey2013b}. We ran chains of $10000$ steps, with $32$ walkers. Model parameters are estimated from the MCMC posterior distributions in the following way: In L band the best-fit values are taken as the mode of the posterior. However, in N-band we use the median for the best-fit. In both bands the uncertainties are taken as the range between the 16th and 84th percentile. The first 1000 steps in the MCMC chain were discarded when calculating the best-fit values and errors.

\section{Results}
\label{sec:res}

\begin{figure*}
   \centering
   \includegraphics[width=\hsize]{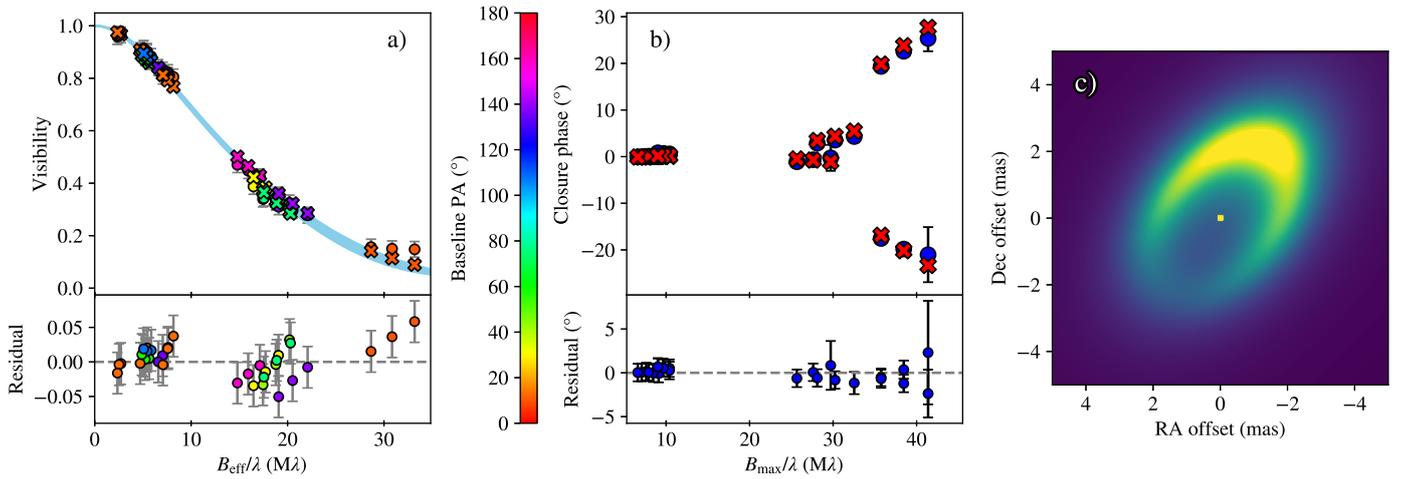}
      \caption{The results of our model fitting for the L-band data with the asymmetric ring model: fits to visibilities (left), fits to closure phases (middle), and the best fit model image at $3.2\ \mu$m (right). Data points are indicated by circles, model values by crosses. On the left panel the symbols are color coded for the baseline position angle. The blue shaded area on the left panel represents the range of model visibility functions taken at different baseline position angles, at $3.2\ \mu$m wavelength.
              }
         \label{fig:model_fit_L_ring}
  \end{figure*}
   
    \begin{figure*}
   \centering
   \includegraphics[width=\hsize]{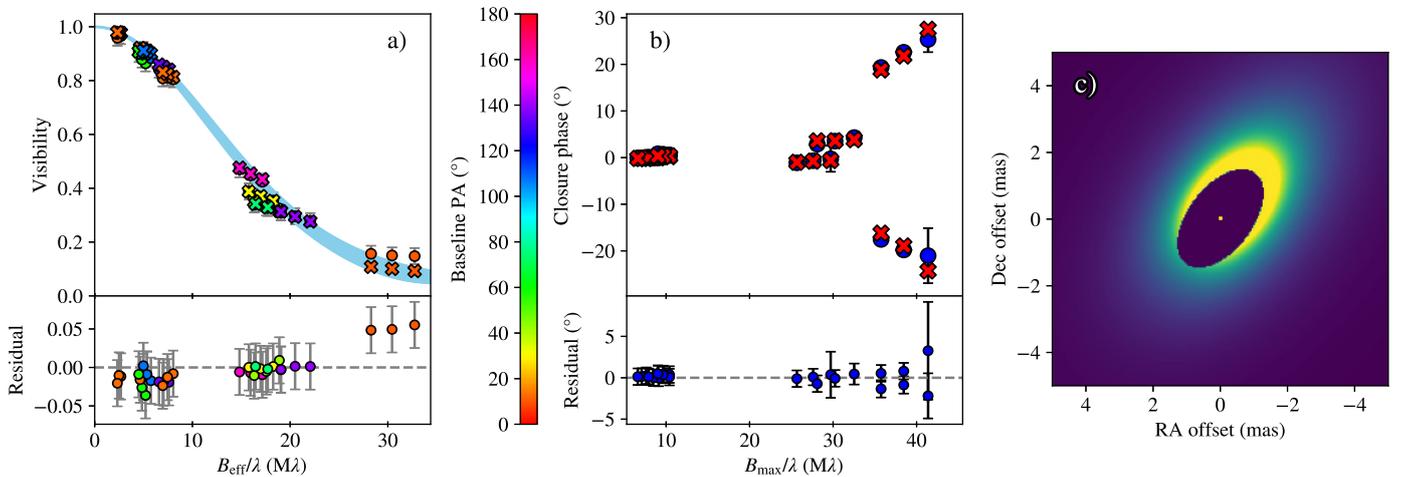}
   \caption{Same as Fig.~\ref{fig:model_fit_L_ring}, but with the flat disk model.
              }
         \label{fig:model_fit_L_flat_temp_grad}
  \end{figure*}
  
\begin{table}
\caption{List of the best-fit parameters, half-light radii, and $\chi^2$-values in our L-band modeling. The $\chi^2$-values are given relative to the number of fitted data points, separately for the visibility and for the closure phase.}
\begin{center}
	\label{tab:res_L}

\begin{tabular}{l c c}
\hline \hline
 & Ring & Flat disk \\
 & model & model \\
\hline
$\theta$ ($\degr$)  &  $141.8^{+1.6}_{-2.4}$  &  $143.0^{+1.9}_{-1.9}$ \\
$\cos\,i$  &  $0.61^{+0.01}_{-0.02}$  &  $0.56^{+0.01}_{-0.01}$ \\
$R_\mathrm{in}$ ($\mathrm{mas}$)  &  $2.71^{+0.05}_{-0.07}$  &  $1.71^{+0.04}_{-0.03}$ \\
$R_\mathrm{in}$ ($\mathrm{au}$)  &  $0.274^{+0.005}_{-0.007}$  &  $0.173^{+0.004}_{-0.003}$ \\
$FWHM_\mathrm{kernel}$ (mas)  &  $5.08^{+0.25}_{-0.09}$  &  \\
$FWHM_\mathrm{kernel}$ (au)  &  $0.51^{+0.03}_{-0.01}$  &  \\
$A_\mathrm{mod}$  &  $0.69^{+0.04}_{-0.05}$  &  $0.54^{+0.03}_{-0.02}$ \\
$\phi_\mathrm{mod}$ ($\degr$)  & $196.3^{+2.0}_{-13.1}$  &  $182.0^{+4.6}_{-6.0}$ \\
$q$  &  & $0.66^{+0.01}_{-0.01}$ \\
\hline
$R_\mathrm{hl}$ (mas)  & $3.28$  &  $3.26$ \\
$R_\mathrm{hl}$ (au)  & $0.33$  &  $0.33$ \\
$\chi^2_V/N_V$ & $0.57$  &  $0.43$ \\
$\chi^2_\mathrm{CP}/N_\mathrm{CP}$ & $0.34$  &  $0.25$ \\
\hline
\end{tabular}

\end{center}
\end{table}

\begin{table*}
\caption{List of the fitted and fixed parameters in our N-band modeling with the 2D Gaussian model. The $\chi^2$ values are given relative to the number of fitted data points ($N = 6$). }
\begin{center}
	\label{tab:res_N}

\begin{tabular}{l c c c c c}
\hline \hline
 & $8.5\ \mu$m & $9.5\ \mu$m & $10.5\ \mu$m & $11.5\ \mu$m & $12.5\ \mu$m \\
\hline
\multicolumn{6}{l}{Fitted parameters}\\
$HWHM_\mathrm{Gaussian}$ ($\mathrm{mas}$)  & $8.79^{+1.84}_{-2.50}$  &  $10.68^{+1.56}_{-1.83}$  &  $12.26^{+1.71}_{-2.07}$  &  $13.21^{+2.29}_{-2.90}$  &  $11.06^{+5.85}_{-6.83}$ \\
$HWHM_\mathrm{Gaussian}$ ($\mathrm{au}$)  & $0.89^{+0.19}_{-0.25}$  &  $1.08^{+0.16}_{-0.19}$  &  $1.24^{+0.17}_{-0.21}$  &  $1.34^{+0.23}_{-0.29}$  &  $1.12^{+0.59}_{-0.69}$ \\
$F_\mathrm{tot}$ ($\mathrm{Jy}$)  & $12.6^{+1.0}_{-0.9}$  &  $20.7^{+1.3}_{-1.2}$  &  $20.8^{+1.3}_{-1.3}$  &  $16.9^{+1.2}_{-1.2}$  &  $11.0^{+1.6}_{-1.2}$ \\
\multicolumn{6}{l}{Fixed parameters}\\
$F_\mathrm{tot,\star}$ ($\mathrm{Jy}$)  & $0.13$ & $0.11$ & $0.09$ & $0.07$ & $0.06$ \\
$\theta$ ($\degr$)  & \multicolumn{5}{c}{ $133.0$ } \\$\cos\,i$ & \multicolumn{5}{c}{ $0.68$ } \\\hline
$\chi^2/N$ & $0.08$  &  $0.20$  &  $0.37$  &  $0.38$  &  $0.07$ \\
\hline
\end{tabular}

\end{center}
\end{table*}

In the L-band we model the visibilites and closure phases from 2019 March and May. We use the spectral aspect in our data by fitting three wavelengths ($3.2,\ 3.45,\ 3.7\ \mu$m) simultaneously. We average the data in a $0.2\ \mu$m wide window around each fitted wavelength. We fit 36 visibility (2 epochs $\times$ 3 wavelengths $\times$ 6 baselines) and 24 closure phase (2 epochs $\times$ 3 wavelengths $\times$ 4 triangles) data points. The resulting fits are presented in Figures~\ref{fig:model_fit_L_ring} and \ref{fig:model_fit_L_flat_temp_grad} for the ring model and flat disk model, respectively. In these figures the left panel shows the visibilities against the deprojected baseline length (in spatial frequency units) in which the effect of the object inclination is removed.
The MCMC posterior distributions are shown in Figures~\ref{fig:cornerplot_L_ring} (ring model) and \ref{fig:cornerplot_L_flat_temp_grad} (flat disk model). Table~\ref{tab:res_L} lists the resulting best-fit parameter values.  In the modeling the stellar flux contribution with respect to the total flux is a fixed parameter. For the stellar flux we use the values $0.81$, $0.71$ and $0.62$~Jy at $3.2$, $3.45$, $3.7$ $\mu$m respectively, based on our spectral energy distribution (SED) modeling. The total flux of the star-disk system from MATISSE data is $10.8 \pm 0.8$~Jy, with very weak wavelength dependence. Using the stellar and total fluxes we calculate a flux ratio for each of the three wavelengths. The other fixed parameters are the luminosity of the central star $L_\star = 16\ \Lsun$, and the distance of the system $d = 101.2$~pc, both required only by the flat disk model.

In order to be able to compare the disk sizes resulting from the different modelings, we introduce a half-light radius ($R_\mathrm{hl}$), a single robust measure of the size of the brightness distribution, as follows:
\begin{equation}
\frac{F_\mathrm{tot}}{2} = \int_{0}^{R_\mathrm{hl}} 2\pi r I_\nu\left(r\right) \mathrm{d}r.
\end{equation}
This is the usual definition used in the IR interferometry literature \citep[e.g,][]{Leinert2004,Varga2018}. We note that the stellar flux is not taken into account in the calculation of $R_\mathrm{hl}$.
In addition to the ring and flat disk models, we also model the L band emission with a Gaussian plus an additional point source model. The results of this model are presented in the Appendix~\ref{sec:gauss_model}.  

All our models fit the data reasonably well. Comparing the L-band models by the $\chi^2$ values, the Gaussian model provides the best fit to the data, although the $\chi^2$ values does not differ much between the models. Looking at the best-fit model images, the ring model and the flat disk model give similar brightness distributions. The images show a strongly asymmetric inclined ring with the brightness maximum located at NW. The corresponding position angle (measured from N towards E) is ${331\degr}_{-12\degr}^{+10\degr}$. In comparison, the additional point source in the Gaussian modeling lies at a similar position angle of $309\degr$. The point source is located at $0.19$~au from the center, which is very similar to the value for the inner radius ($0.17$~au) in the flat disk model. In the ring and flat disk models the ratio of the brightest to faintest surface brightness in the azimuthal variation is $3.5 \pm 0.2$. In the Gaussian modeling the flux ratio of the additional point source is $0.09 \pm 0.01$, comparable to the flux contribution of the central star. All three models agree well on the basic geometric parameters: the inclination is ${52\degr}^{+5\degr}_{-7\degr}$, the disk position angle is $143\degr \pm 3\degr$, and the half-light radius of the L-band emitting region is $0.33 \pm 0.01$~au. We note that due to the sparse $uv$-coverage, typical for IR interferometric observations, multiple models with different brightness distributions may also fit the data. 

In the N-band, we model the correlated fluxes from 2019 March with the 2D Gaussian model. Five wavelengths between $8.5$ and $12.5\ \mu$m are fitted independently. Correlated fluxes are averaged in a $1\ \mu$m wide window around each fitted wavelength. There are 6 fitted data points per wavelength bin. The N-band fit results are shown in Table~\ref{tab:res_N}. The half-radii are between $0.89$ and $1.34$~au, showing an increasing trend with wavelength from $8.5\ \mu$m to $11.5\ \mu$m. The N-band emitting region is $3-4$ times larger than the L-band one.

We checked the consistency of our modeling between the two MATISSE bands by extrapolating the best-fit L-band flat disk model to the N-band. The resulting model image calculated at $10.5\ \mu$m has a half-light radius of $11.0$~mas ($1.1$~au), which is very close to the value we got from the Gaussian fit to the N-band data ($12.3$~mas). This result indicates that the flat-disk model is capable to broadly represent the disk emission in both MATISSE bands. We also extrapolated the flat disk model to the near-IR, in order to make comparisons with PIONIER and GRAVITY observations, discussed in the next section.

\section{Discussion}
\label{sec:discussion}

As mentioned in Sect.~\ref{sec:intro}, discrepancies in the findings of near-IR interferometric studies  prevent us from drawing a consistent picture of the inner disk structure of HD 163296. The L-band MATISSE data-set is consistent both with a ring-like and with a centrally peaked geometry. The PIONIER model image of \citet{Lazareff2017}  and the new GRAVITY reconstructed images (GRAVITY Collaboration, in prep.) show an asymmetric ring. However, the recent directly reconstructed PIONIER image \citep{Kluska2020} is centrally peaked. \citet{Kluska2020} performed an image reconstruction simulation on the  best-fit ring model of \citet{Lazareff2017}, by calculating synthetic interferometric observables from the model image. They found that the reconstructed image on the synthetic data-set is in agreement with the reconstructed image on the real data-set. This indicates that the presence of a central cavity is plausible,  and the ambiguity in the image reconstruction may be caused by the insufficient resolution. In a K-band interferometric study of a sample of Herbig Ae/Be stars, including HD 163296, a ring model based on \citet{Lazareff2017} was fitted \citep{Perraut2019}. The best-fitting model to HD 163296 is dominated by the centrally peaked component. Similarly, \citet{Setterholm2018} found that a simple Gaussian fits better than a thin ring or uniform disk to H- and K-bands interferometric data. It is important to note that neither \citet{Perraut2019} nor \citet{Setterholm2018} modeled closure phases, so their models are symmetric. Further interferometric observations, with optimized $uv$-coverage at the longest possible baselines, are needed to properly the resolve the disk structure within $r=0.3$~au. VLTI currently provides baseline lengths up to $\sim 130$~m, but there are ongoing efforts to open the longest AT baseline (220 m). Having such a long baseline could provide highly valuable constraints on the inner disk structure.

Our results  for the disk inclination and position angle are in line with earlier near-IR interferometric observations ($i = 45^\circ-50^\circ$, $\theta = 126^\circ-134^\circ$), with ALMA observations ($i = 47^\circ$, $\theta = 133^\circ$),  and with SPHERE observations ($\theta = 134\degr-137.5\degr$) \citep{Lazareff2017,Kluska2020,Perraut2019,Setterholm2018,Huang2018,Isella2018,Muro-Arena2018}. Thus, we conclude that the inner disk of HD 163296 is well aligned with the outer disk.

\subsection{The nature of the asymmetry}

\begin{figure}
   \centering
   \includegraphics[width=\hsize,trim=0 0 0 0, clip]{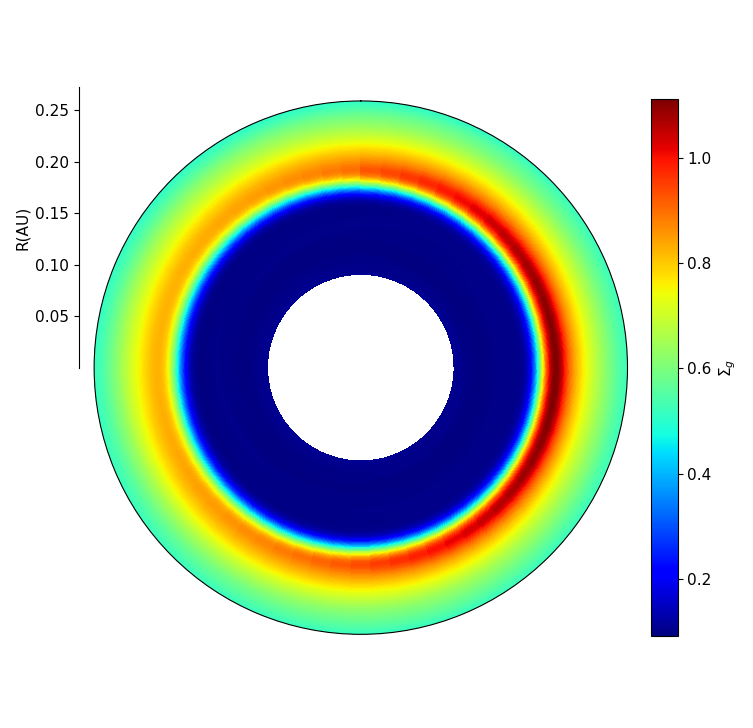}
      \caption{Gas surface density at the inner edge of a disk obtained with numerical simulation. The asymmetry in density is due to a vortex formed by the Rossby wave instability.
              }
         \label{fig:simu_density}
\end{figure}

A circumstellar disk seen at an inclined viewing angle can show brightness asymmetries even if the structure of the disk is perfectly circularly symmetric. These asymmetries arise from radiative transfer effects. An example is the anisotropic scattering of the stellar light \citep[e.g.,][]{Pinte2009}. In the case of HD 163296, the star contributes less than $7\%$ to the total L-band flux, thus scattered light is negligible in the disk emission. A further inclination effect is that the far side of the inner rim appears brighter than the (self-shadowed) near side\footnote{In the disk of HD 163296, the near side of the rim is located towards NE. This is supported by the velocity measurements of the jet and associated Herbig-Haro objects \citep{Devine2000}, and by the velocity mapping of the molecular gas emission \citep{Teague2019}.} \citep[e.g.,][]{Isella2005}. In both scenarios the brightness maximum should correspond to the position angle of the minor axis of the projected image. Surprisingly, our MATISSE results show that the maximum brightness in the inner disk of HD 163296 is towards the semimajor axis to the NW. This is not compatible with inclination effects, thus, we argue that the asymmetry is caused by an azimuthal variation in the disk structure, or in the dust properties (like grain size).

In the PIONIER model image of \citet{Lazareff2017} the position angle of the brightness maximum is $273\degr$ (measured from N towards E). This is $65\degr$ away from the position angle we get from MATISSE data (${338\degr}^{+3\degr}_{-14\degr}$), using the asymmetric ring model which was also used for the PIONIER modeling. Furthermore, GRAVITY Collaboration (in prep.) report that the position angle of the bright arc has changed from $\approx$$60\degr$ to $\approx$$240\degr$ between 2018 July to 2019 July. All evidence indicate a time-variable morphology of the $r<0.3$~au disk region. Detecting the variable morphology from MATISSE data alone is challenging, because the data-sets taken on the short baselines (three out of the four observations) are barely sensitive to the asymmetries, as the closure phases are within $\pm 1\degr$. Nevertheless, we tried to detect the change in the position of the asymmetry between the 2019 March and May MATISSE epochs, by fitting the corresponding data-sets independently. We used the flat disk model by fixing all parameters to their best-fit values, except for $A_\mathrm{mod}$ and $\phi_\mathrm{mod}$. For 2019 March we get $\phi_\mathrm{mod} = {196\degr}^{+81\degr}_{-60\degr}$ (measured from the major axis), and for 2019 May $\phi_\mathrm{mod} = 179.5\degr \pm 2\degr$. The difference is not significant. We estimate that the orbital periods at the observed radius range ($0.18-0.33$~au) of the ring range from $20$ to $50$ days. The temporal coverage of the interferometric data  (from MATISSE, PIONIER,  and GRAVITY) does not allow us to constrain the rotation period. 

The physical origin of the time-variable disk structure is unclear. Several kinds of magnetohydrodynamic instabilities \citep[e.g.,][]{Flock2015,Flock2017}, Rossby wave instability \citep{Lovelace1999,Meheut2010}, or gravitational instability \citep[e.g.,][]{Durisen2007,Kratter2016} can produce such asymmetric disk features. Alternatively, a (sub-)stellar companion \citep[e.g.,][]{Brunngraber2018,Szulagyi2019} could also be the direct cause of the asymmetry. Our Gaussian plus additional point source model can provide constraints on the flux of such a putative companion. The offset point source in this model has an L-band flux of $0.9$~Jy. We also fitted a physically more realistic model which has two point sources (the central star and the companion), and a circumbinary ring. The ring geometry is the same that we used in the smoothed ring model, except that the circumbinary ring is symmetric. This model predicts an even higher flux for the companion ($1.5$~Jy). These results imply that the companion would be brighter in the L-band than the central star which has a flux of $\sim 0.7$~Jy. The separation of the companion from the central star in the circumbinary ring model is $\sim 0.15$~au. In order to get constraints on the near-IR flux of the putative companion, we fitted the PIONIER data from \citet{Lazareff2017} with the Gaussian plus additional point source model, and also with the circumbinary ring model. The H-band fluxes we get for the companion are in the range of $0.2-0.4$~Jy, assuming a total H-band flux for the system of $6.4$~Jy, and a H-band stellar flux of $2.5$~Jy. 

If the object causing the asymmetry is a stellar companion, its near-IR flux is very likely to be higher than its L-band flux. Our interferometric modeling suggests that this is not the case. If we assume that the companion is a planet, its close location to the star, in addition to its high flux, would classify it as a hot Jupiter. Hot-start planet models \citep[e.g.,][]{Spiegel2012} predict that a young planet with a mass of $10\ \mathrm{M}_{\jupiter}$ may have an effective temperature around $2500$~K and a radius of $2.3\ \mathrm{R}_{\jupiter}$. The IR flux from such planet itself would be too small (a few mJy in L-band) to explain our observations. We also consider that the IR emission originates from a circumplanetary disk. If we assume optically thick thermal dust emission (with a dust temperature of $1500$~K), the required L-band flux could be provided by a disk with a diameter of $\sim 0.1$~au. However, the Hill-sphere diameter of a $10\ \mathrm{M}_{\jupiter}$ mass planet located at $r=0.15$~au is $0.035$~au (assuming a stellar mass of $1.9\ \Msun$), too small to accommodate a circumplanetary disk of the required size. Although a much hotter circumplanetary material filling the Hill-sphere might provide enough L-band flux to match the MATISSE observations, its near-IR flux contribution would be much larger and thus incompatible with the PIONIER data. Therefore, we argue that the observed brightness asymmetry is caused by an asymmetric structure in the disk itself. 

Hydrodynamic instabilities may be responsible for the disk asymmetry we see in our data. In order to test the plausibility of the Rossby wave instability hypothesis, we have performed hydro-dynamical numerical simulations with \textsc{AMRVAC}\footnote{\url{amrvac.org}}\citep{XTE18} of the inner edge of the disk (see also \citealp{RMM20} for the setup). Only the dynamical evolution of the gas is modeled and the dust is considered to follow the gas. This is valid for small dust grain with Stokes number $\mathrm{St} \sim \frac{s\rho_d}{\Sigma_g}<<1$, where $s$ is the size of the dust grain, $\rho_d$ its internal density, and $\Sigma_g$ the gas surface density. The strong density gradient at the inner edge of the disk creates a minimum in the vortensity (or potential vorticity) profile. This is unstable due to the Rossby wave instability that form vortices \citep{Lovelace1999}. An inverse cascade in this 2D simulation is eventually responsible for the survival of a unique large scale vortex, a similar evolution would be obtained within a thin 3D disk \citep{MKC12}. This anticyclone is a high pressure and density region, as can be seen on Fig.~\ref{fig:simu_density}, which is known to concentrate dust \citep{BAR95}. Although neither the gas nor the vortex rotate at the local Keplerian frequency, in both cases, the difference from Keplerian rotation is small. Thus, Keplerian motion is a good approximation for the variability time scale in the vortex scenario. For further details on the dynamical signatures of vortices we refer to \citep{Robert2020}.

Density enhancement alone is not expected to increase the surface brightness of the vortex region, assuming optically thick emission. What is required to produce increased radiation, is a change in the dust properties. A potential mechanism for that is the production of small grains by grain collisions which also increase the local temperature. The combined effect of more small grains and increased temperature will be a local increase of surface brightness. Such mechanism could be an explanation for the asymmetric feature in HD 163296. The azimuthal extent of the vortex roughly corresponds to that of the asymmetry in our ring and flat disk model images, although in our modeling we only applied first-order azimuthal modulation. To fully test this hypothesis, a hydro-dynamical simulation including the dust dynamics is needed, as the Stokes number of solids of a given size strongly varies at the edge of a disk due to the gas surface density profile. This hydro-dynamical simulation should then be coupled to radiative transfer code to provide synthetic images and synthetic interferometric observables.

\citet{Rich2020} report Hubble Space Telescope (HST) coronagraphic imaging of HD 163296, and multiwavelength photometric monitoring from 2016-2018. They found azimuthally asymmetric surface brightness variations between the HST epochs, and they conclude that the disk illumination varies on $<3$ months time scales. They suggest that the origin of the brightness variations is shadowing by a structure residing within $0.5$~au radius. This result fits well in our picture of the variable inner disk morphology, and it may also support our vortex scenario. In previous 3D vortex simulations it was found that the disk is vertically more extended at the location of the vortices \citep{MeheutMeliani2012}. Thus, a vortex can indeed be a source of shadowing.

To shed more light on the origin of the asymmetry, it would be highly important to measure its rotation period. For this, monitoring observations with a few days to a week cadence are needed. Coordinated observations using near- and mid-IR interferometric instruments (e.g., GRAVITY and MATISSE) quasi-simultaneously are also desired in order to reveal the chromatic nature of the asymmetry, and to improve the quality of the reconstructed images \citep{Sanchez2018}. The assessment of the chromaticity can help to constrain the physical nature of the asymmetry.

\subsection{The size of the IR emitting region}
\label{sec:hlr}

\begin{figure}
   \centering
   \includegraphics[width=\hsize]{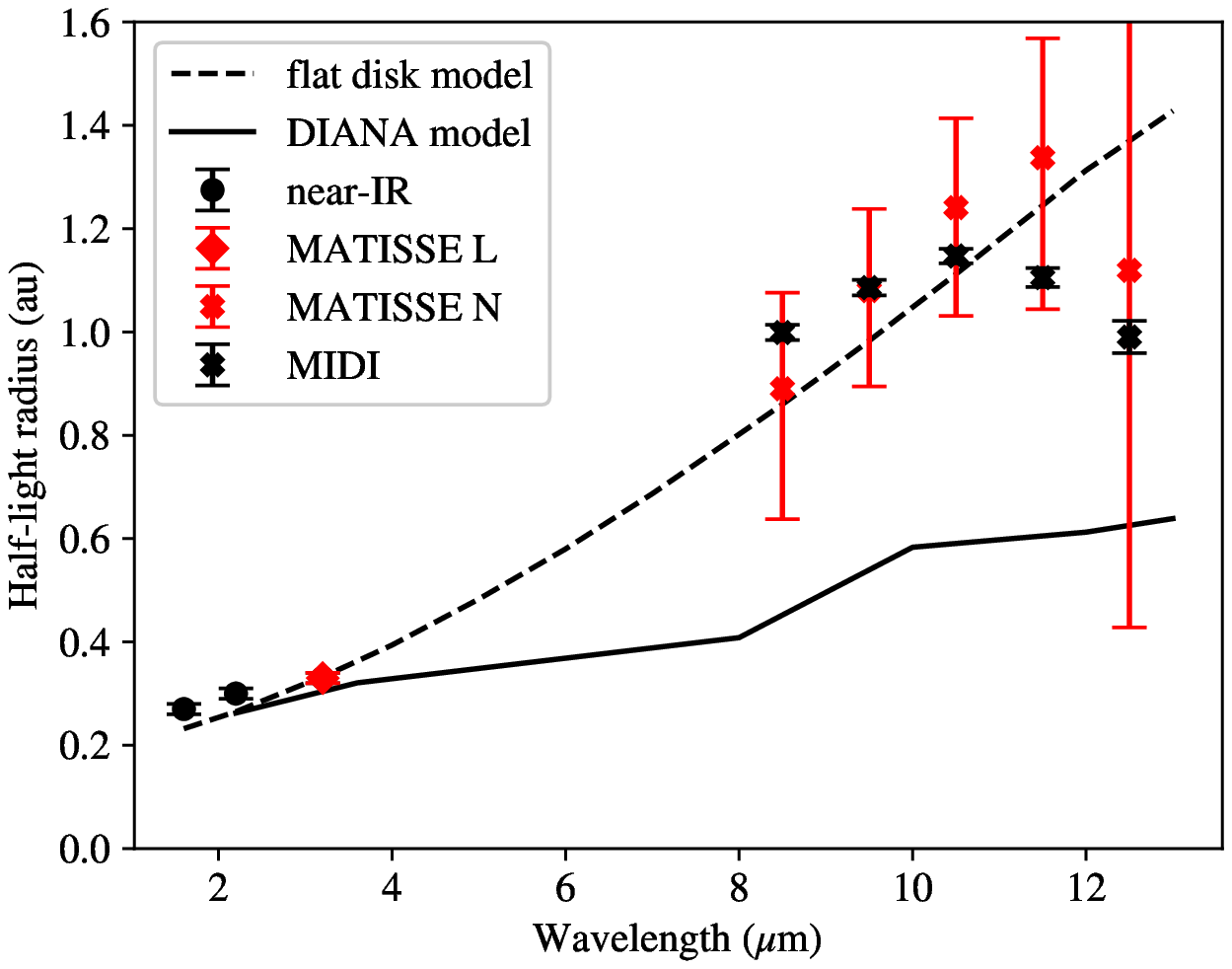}
      \caption{Characteristic size of the emitting region as function of wavelength. The red points are derived from our MATISSE data. The $1.6$ and $2.2\ \mu$m points come from PIONIER \citep{Lazareff2017,Kluska2020} and GRAVITY  \citep{Perraut2019} observations, respectively. Black crosses represent our new fits to MIDI data published by \citet{Varga2018}.
      The solid line is from the DIANA radiative transfer model of HD 163296 \citep{Woitke2019}. The dashed line corresponds to our flat disk model.
              }
         \label{fig:hlr}
  \end{figure}
  
In Fig.~\ref{fig:hlr} we show the half-light radius of the disk emitting region derived from near- and mid-IR interferometric observations as function of wavelength. The new L-band size from our observations is very close to the K-band size, suggesting that the near-IR and the $3-4\ \mu$m emissions originate from the same disk region. There is a general increasing trend of the half-light radius with wavelength.  This is because towards longer wavelengths we become sensitive to material at lower temperatures, located further from the central star. In the N-band the characteristic size of the disk emission as traced with our MATISSE data (red crosses) and archival MIDI data (black crosses) is much larger than at shorter wavelengths. Furthermore, the apparent size in the 10~$\mu$m silicate feature is somewhat larger than in the adjacent continuum. This behavior is expected based on radiative transfer calculations \citep{vanBoekel2005_flaring}: the emission in the N-band arises in part from the warm surface layer and in part from the cooler (and hence at a given wavelength apparently more compact) disk interior. In the N-band the warm surface layer is optically thin in the vertical direction. Therefore the relative contribution of the spatially more extended surface layer emission follows its opacity curve, explaining both why we see an emission feature and why the spatial extent of the emission is highest in the emission band.

In Sect.~\ref{sec:res} we demonstrated that our best-fitting flat disk model to the L-band data is capable of reproducing the observed increase in size between L- and N-band. We note however, that this model does not include the effects of the silicate opacity. As it can be seen in Fig.~\ref{fig:hlr}, the flat disk model (represented by the dashed line) is also in good agreement with the disk sizes measured by PIONIER and GRAVITY in the near-IR. We note that the H- and K-band sizes found by \citet{Setterholm2018} were the same within errors, with the longer CHARA baselines.

To compare our N-band results with older MIDI observations, we fitted the same 2D Gaussian models that we used for MATISSE data, to MIDI data taken from \citet{Varga2018}. The resulting N-band disk sizes from MIDI are plotted in Fig.~\ref{fig:hlr} (black crosses). The N-band sizes derived from MATISSE and MIDI observations are in good agreement. The error bars on the MATISSE N-band data points are much larger than those on the MIDI points. We attribute this to the following causes: 1) The MIDI data-set has many more data points (27 MIDI vs. 6 MATISSE correlated flux points). 2) 
The MIDI data-set samples a larger range of spatial scales ($1\ \mathrm{au}<\vartheta<9\ \mathrm{au}$), compared to MATISSE ($3\ \mathrm{au}<\vartheta<10\ \mathrm{au}$). These two factors enable more precise size measurements for the MIDI data.

For comparison, we overplot a size-wavelength curve from a radiative transfer modeling of HD 163296 on Fig.~\ref{fig:hlr}. The disk model is taken from the DIANA project website\footnote{\url{https://www.univie.ac.at/diana/index.php/user/user/viewuser}} \citep{Woitke2019}. We used the radiative transfer code MCMax \citep{Min2009} to generate images of the disk at different wavelengths. We calculated the half-light radius including the stellar flux contribution, and used circular apertures. The half-light radii of the DIANA disk model are in agreement with H-, K-, and L-band observations, but the observed N-band sizes are significantly larger than the model values. A possible explanation for this difference is that the dust grain size distribution changes with the distance from the star, i.e., that the innermost disk lacks small ($<2-5\ \mu$m) silicate grains, while a bit further away (at $\sim$$1$~au) small grains are still abundant. \citet{Millan-Gabet2016} found similar results for a number of Herbig Ae/Be stars, by using a 2-rim model which provided a better fit to IR interferometric and SED-data, compared to an inner rim plus flared disk geometry. Their model featured an inner rim with large grains, and an outer rim with larger scale height containing smaller grains. The inferred inner rim radius for HD 163296 ($0.39$~au) seems to be incompatible with our L-band half-light radius, but their outer rim radius ($1.1$~au) matches well our N-band half-light radii. 

A further evidence for this scenario might come from interferometric modeling. We fitted the MIDI data from \citet{Varga2018} with the flat disk temperature gradient model at three wavelengths ($8$, $10.7$, and $13\ \mu$m). The fits at $8$ and $10.7\ \mu$m, representing the continuum emission, are consistent with a continuous disk beginning at the dust sublimation radius. While at $10.7\ \mu$m which is in the middle of the silicate feature, the data is better represented with a disk with an inner hole (with $\sim$$0.6$~au inner radius).  
Thus, it is likely that the structure of the N-band continuum emission is continuous, while the silicate emission comes from a larger region exhibiting an inner gap. A strong N-band silicate feature is only expected from small ($<1\ \mu$m) silicate grains. Thus, a radial change in the grain size distribution, probably due to grain growth in the innermost disk, naturally explains our findings. This scenario was found to be the case for the transitional disk T Cha \citep{Olofsson2013}. 

An alternative solution for the larger than expected N-band sizes could be the presence of dust in a halo-like structure. \citet{Ellerbroek2014} studied optical to near-IR spectra of the jet of HD 163296. They detected fadings in the optical which coincided with near-IR brightenings. They propose that this can be explained by dust lifted high above the disk plane, in a disk wind launched at $\gtrsim 0.5$~au radii. Such optically thin dust cloud above the disk might also be consistent with our results. While our analysis with simple geometric models provides valuable insights into the N-band disk structure, to reach a consistent multiwavelength view on the dust distribution, more detailed modeling is needed.

\subsection{The nature of the dust sublimation zone}

The dust sublimation zone is not an abrupt boundary between the innermost gaseous disk and the dusty disk regions, but rather a continuous transition region \citep[e.g.,][]{Isella2005,Kama2009,Dullemond2010}. The location and properties of this region heavily depend on the grain sizes and composition. 
To estimate a radial range for the dust sublimation zone of HD 163296, we apply Eq.~9 from \citet{Dullemond2010}. This equation contains the factor $\epsilon$ which is expressed by dividing the effectiveness of the emission at the wavelength at which the dust radiates away the heat with that of the absorption at the wavelength of the stellar emission. It represents how efficiently the dust cools. We assume a sublimation temperature of $1500$~K, a stellar luminosity of $16\ \Lsun$, and a range for the cooling efficiency factor of $0.1 < \epsilon < 1.0$. This yields a wide radial range of $0.14-0.43$~au for the sublimation zone. Our flat disk model features a sharp inner rim, however, this is likely not a physically realistic representation of the sublimation zone. Due to the sparse $uv$-coverage and insufficient resolution, we cannot recover the radial brightness profile of the dust sublimation zone from the data. Nevertheless, our modeling results may give some hints on the location. The radius of the inner rim in our flat disk model is $0.18$~au, close to the lower limit of the estimated radial range for the sublimation zone. Furthermore, in our ring model the disk surface brightness near the star is still $20\%$ of the value at the brightness maximum. Both models suggest that non-negligible fraction of the mid-IR light comes from a region where $\mu$m-sized grains  may not survive. Considering Eq. 9 of \citet{Dullemond2010}, we consider the following solutions for this issue. The first is the presence of large ($\gtrsim 10\ \mu$m) dust grains which have a high cooling efficiency factor ($\epsilon \approx 1$). At $0.18$~au radius these large grains would have a temperature of only $1300$~K. The second solution is the presence of small grains made of refractory materials which have a higher sublimation temperature. A refractory grain with $\epsilon \approx 0.1$ at $0.18$~au should endure a temperature of $2300$~K. Additionally, gas emission within the dust sublimation radius might contribute to the near- and mid-IR radiation. 
 
In the following we discuss these three scenarios in the following order: 
1) gas emission, 2) refractory grains, 3) large grains. Several studies proposed that gas continuum emission inside the dust sublimation radius can provide significant contribution to the near-IR radiation of protoplanetary disks \citep[e.g.,][]{Muzerolle2004,Eisner2007temp,Isella2008,Weigelt2011}. \citet{Tannirkulam2008} successfully modeled K-band interferometric data of HD 163296 by using a rim model with and additional uniform disk component in the inner cavity, attributed to gas continuum emission.
\citet{Benisty2010_HD163296} performed radiative transfer modeling of HD 163296 using AMBER H- and K-bands observations, and they found a good fit to the data with a model where the disk emission originates from a region between $0.1$ and $0.45$~au. They tested an accretion disk model with gas in local thermodynamic equilibrium (LTE), a non-LTE model of thin layers of gas in the disk atmosphere, and a model with hot gas ($T = 8000$~K). However, they found that these gas models are inconsistent with observational data. 

\citet{Benisty2010_HD163296} propose instead the presence of refractory dust grains in the $0.1-0.45$~au region. How dust can survive the high temperatures ($2100-2300$~K) inside, is not well established. From the dust types they discuss, iron seems to be a good candidate. However, thermal equilibrium calculations show that above $1500$~K only Al and Ca bearing minerals persist, and Fe containing grains are not stable any more \citep[e.g.,][]{Scott2007}. One of the most refractory mineral known is corundum (Al$_2$O$_3$), which is stable up to $\sim$$1800$~K. The existence of refractory dust in the disk of HD 163296, sublimating at $\sim$$1850$~K, was also proposed by \citet{Tannirkulam2008}. As an additional scenario, we consider the presence of large dust grains in the inner disk. Large silicate grains have a large cooling efficiency, so they can survive at a distance from the star similar to small refractory grains. The depletion of small grains in the inner disk are also supported by our findings regarding the structure of the N-band emitting region, discussed in Sect.~\ref{sec:hlr}. Testing the plausibility of these scenarios needs more detailed modeling, which is not in the scope of this paper. 

\section{Summary}
\label{sec:summary}

In this study we have presented the first MATISSE observations of the disk around the Herbig Ae/Be star HD 163296. The object is resolved both in L- and N-bands. The L-band closure phases indicate significant brightness asymmetry. We modeled the disk using various geometric models, including an asymmetric ring, an asymmetric flat disk with inner cavity, and a 2D Gaussian. All three geometries were used to model the L-band disk structure, while only the last was fitted to the N-band data. Our main findings are as follows:

   \begin{enumerate}
    \item Our models can describe well the L-band visibilities and closure phases. The half-light radius of the L-band emitting region is $0.33\pm 0.01$~au, the inclination is ${52\degr}^{+5\degr}_{-7\degr}$, and the position angle is $143\degr \pm 3\degr$. 
    
    \item The N-band emitting region has a half-light radius of $0.9-1.3$~au, showing an increasing trend with wavelength from $8.5\ \mu$m to $11.5\ \mu$m. The observed N-band sizes are significantly larger than the prediction from a radiative transfer model. A possible explanation for this difference is the lack of small silicate grains in the inner disk regions ($r \lesssim 0.6$~au).
        
    \item The size of the L-band emitting region is very similar to the near-IR sizes, and $3-4$ times smaller than the N-band size. This suggests that the same emission component dominates the disk emission from the near-IR wavelengths to the L-band.
    
    \item There is no significant misalignment of the L-band emitting disk region, with respect to near-IR and mm (ALMA) measurements.
    
    \item Our modeling reveals a significant brightness asymmetry in the L-band disk emission. The brightness maximum of the asymmetry is located at the NW part of the disk image, nearly at the position angle of the semimajor axis. The position of the brightness asymmetry suggests that it is caused by a variation in the disk structure in or near the inner rim. 
    
    \item Comparing our result on the location of the asymmetry to PIONIER \citep{Lazareff2017} and GRAVITY (GRAVITY Collaboration, in prep.) results, we find that the morphology of the $r<0.3$~au disk region is time-variable. We propose that the asymmetric structure orbits the star with a period of $\sim$$20-50$~days.
    
    \item The physical origin of the rotating asymmetry is unclear. We tested a hypothesis where a vortex created by Rossby wave instability causes the asymmetry. We find that a unique large scale vortex may be compatible with our data. Further hydro-dynamical simulations, and radiative transfer modeling are needed to fully evaluate this scenario. 

    \item Our models predict that a non-negligible fraction of the L-band disk emission originates inside the dust sublimation radius for $\mu$m-sized grains. For the origin of this emission, we consider the presence of refractory grains  and large ($\gtrsim 10\ \mu$m-sized) grains.
      
   \end{enumerate}

\begin{acknowledgements}
MATISSE was designed, funded and built in close collaboration with ESO, by a consortium composed of institutes in France (J.-L. Lagrange Laboratory -- INSU-CNRS -- C\^ote d’Azur Observatory -- University of C\^ote d'Azur), Germany (MPIA, MPIfR and University of Kiel), the Netherlands (NOVA and University of Leiden), and Austria (University of Vienna). The Konkoly Observatory and Cologne University have also provided some support in the manufacture of the instrument.

This research has made use of the services of the ESO Science Archive Facility.

The research of J. Varga and M. Hogerheijde is supported by NOVA, the Netherlands Research School for Astronomy.

T. Henning acknowledges support from the European Research Council under the Horizon 2020 Framework Program via the ERC Advanced Grant Origins 83 24 28.

A. Gallenne acknowledges support from the European Research Council (ERC) under the European Union’s Horizon 2020 research and innovation programme under grant agreement No 695099 (project CepBin).

P. \'Abrah\'am acknowledges support from the Hungarian NKFIH OTKA grant K132406, and from the European Research Council (ERC) under the European Union's Horizon 2020 research and innovation programme under grant agreement No 716155 (SACCRED).

We all would like to thank our colleagues and friends, Olivier Chesneau, Michel Dugu\'e, J. Alonso, A. Glazenborg, H. Hanenburg, J. Idserda, T. Phan Duc, and K. Shabun for their contributions to MATISSE.

\end{acknowledgements}

\bibliographystyle{aa}
\bibliography{ref_MIDI_atlas}

\begin{appendix}

\section{Data processing flow chart}
In fig.~\ref{fig:flowchart} we present a flow chart on our data processing workflow.

\tikzstyle{decision} = [diamond, draw,
    text width=6.0em, text badly centered, node distance=2.5cm, inner sep=0pt]
\tikzstyle{block} = [rectangle, draw, text centered, rounded corners, text width=13em, minimum height=4em]

\tikzstyle{data} = [ 
        trapezium,
        trapezium left angle=70,
        trapezium right angle=110,
        text width=3cm,
        inner ysep=5pt, 
        minimum width=4.4cm, 
        text centered,
        draw=black
    ]

\tikzstyle{bigdata} = [ 
        trapezium,
        trapezium left angle=70,
        trapezium right angle=110,
        text width=5.0cm,
        inner ysep=5pt, 
        minimum width=5.0cm, 
        text centered,
        draw=black
    ]
\tikzstyle{line} = [draw, very thick, color=black, -latex']
\tikzstyle{cloud} = [draw, ellipse,fill=red!20, node distance=2.5cm,
    minimum height=2em]

 \begin{figure*}
   \centering
\begin{tikzpicture}[scale=2, node distance = 2cm, auto]

    \node [data] (init) {\textbf{Raw data}\\ \begin{tabular}{c | c }
    L & N
    \end{tabular}};

    \node [block, below of=init] (datared) {\textbf{DRS data reduction}\\
    \begin{tabular}{c | c }
    incoherent & coherent
    \end{tabular}};
    
    \node [data, below of=datared] (reddata) {\textbf{Reduced data}\\
    \begin{tabular}{c | c }
    $V,\ CP$ & $F_\mathrm{corr},\ CP$\\ 
    \end{tabular}};
    
    \node [block, below of=reddata] (calibration) {\textbf{Calibration}\\
    \begin{tabular}{c | c }
    visibility & flux\\ 
    calibration & calibration\\
    \end{tabular}};
    
    \node [data, left of=reddata,node distance=4.5cm] (calibreddata) {\textbf{Reduced\\ calibrator data}\\
    \begin{tabular}{c | c }
    $V,\ CP$ & $F_\mathrm{corr},\ CP$\\ 
    \end{tabular}};
    
    \node [data, right of=reddata,node distance=4.5cm] (calibspec) {\textbf{Calibrator spectrum}\\
    (only for N-band)};
    
    \node [data, right of=calibration,node distance=5.5cm] (calibdiam) {\textbf{Calibrator diameter}};
    
    \node [data, below of=calibration] (calibdata) {\textbf{Calibrated data}\\
    one file per exposure};
    
    \node [block, below of=calibdata] (avg) {\textbf{Averaging exposures}};
    
    \node [block, left of=avg,node distance=5.5cm] (error) {\textbf{Error analysis}};
    
    \node [decision, below of=avg] (filter) {\textbf{Accept / reject data}};
    
    \node [bigdata, below of=filter,yshift=-1cm] (finaldata) {\textbf{Final data}\\
    Calibrated, averaged\\
    \begin{tabular}{c | c }
    $V,\ CP$ & $F_\mathrm{corr},\ CP$\\ 
    2019 March, May & 2019 March \\
    \end{tabular}};
    
    \node [bigdata, right of=filter,node distance = 4.7 cm] (rejected) {\textbf{Rejected data}\\
    \begin{tabular}{c | c }
    2019 June & 2019 May, June \\
    \end{tabular}};
    
    \path [line] (init) -- (datared);
    \path [line] (datared) -- (reddata);
    \path [line] (reddata) -- (calibration);
    \path [line] (calibreddata) |- (calibration);
    \path [line] (calibspec) -- (calibration);
    \path [line] (calibdiam) -- (calibration);
    \path [line] (calibration) -- (calibdata);
    \path [line] (calibdata) -- (avg);
    \path [line] (avg) -- (filter);
    \path [line] (calibdata) -| (error);
    \path [line] (error) |- (filter);
    \path [line] (filter) -- (finaldata);
    \path [line] (filter) -- (rejected);

\end{tikzpicture}
      \caption{Flow chart of our data processing workflow. $V$ indicates visibility, $CP$ indicates closure phase, and $F_\mathrm{corr}$ indicates correlated flux. }
         \label{fig:flowchart}
\end{figure*}
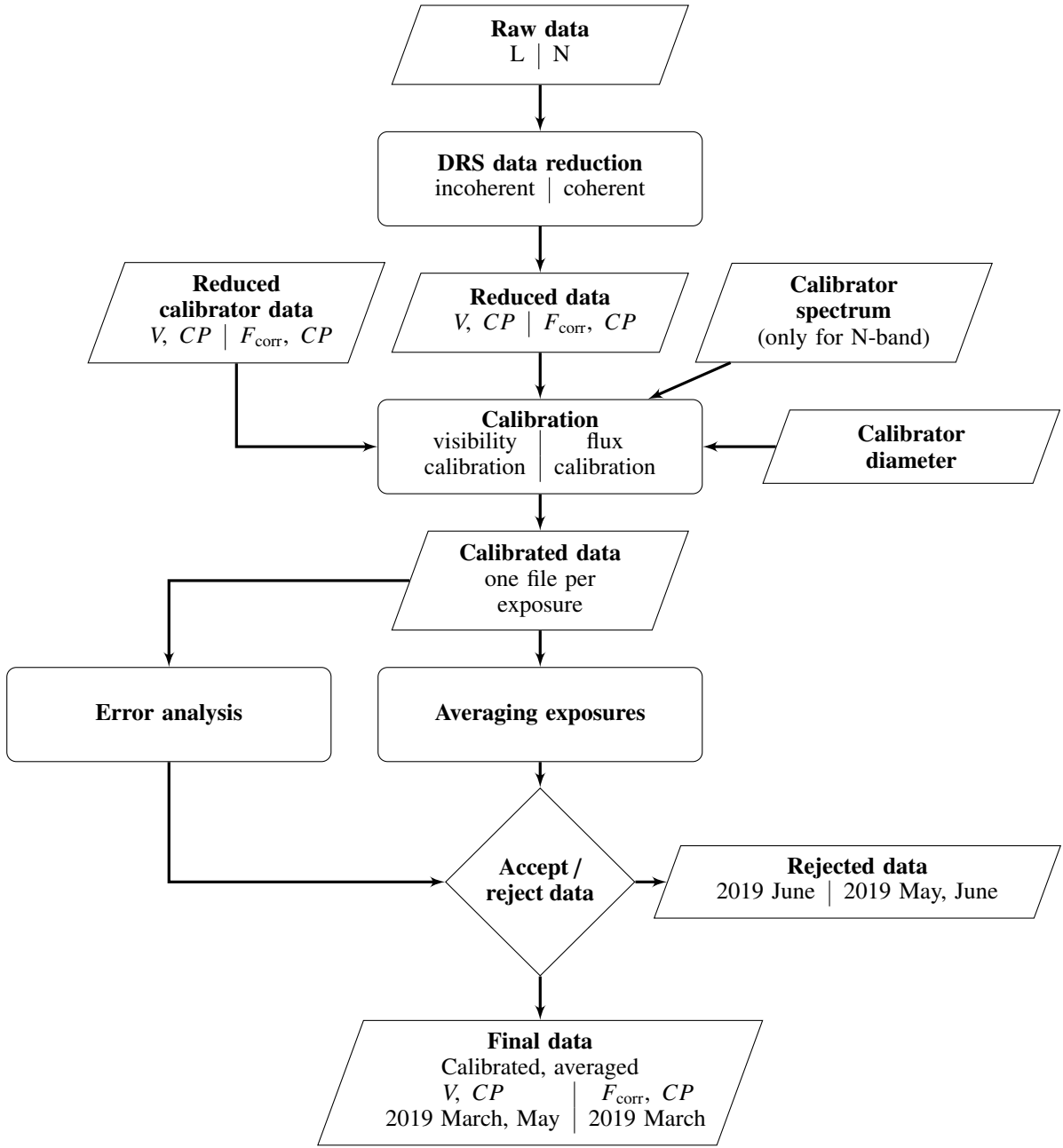

   \section{Error analysis}
   \label{sec:error}
       	\begin{table}
       	
		\caption{Error statistics of our observations: $\sigma_\mathrm{r}$ is the random noise between the spectral channels, $\sigma_\mathrm{sys, s}$ is the short-term systematic uncertainty, and $\sigma_\mathrm{sys, l}$ long-term systematic uncertainty. For visibilities absolute, but for correlated fluxes relative errors are given. The methods to estimate the uncertainties are explained in the Appendix~\ref{sec:error}.}
		
		\small
		\begin{center}
			\label{tab:error}
			\begin{tabular}{l c c c c c c }
				\hline
				\hline
				\multicolumn{7}{c}{L-band}\\
				Date & \multicolumn{3}{c}{Visibility} & \multicolumn{3}{c}{Closure phase ($^\circ$)} \\
				& $\sigma_\mathrm{r}$ & $\sigma_\mathrm{sys, s}$ & $\sigma_\mathrm{sys, l}$ & $\sigma_\mathrm{r}$ & $\sigma_\mathrm{sys, s}$ & $\sigma_\mathrm{sys, l}$  \\
				\hline
				2019-03-23 & 0.002 & 0.013 & 0.017 & 0.13 & 0.20 & 0.05\\
2019-05-06 & 0.001 & 0.004 & 0.011 & 0.38 & 0.87 & 0.09\\
2019-06-26 & 0.004 & 0.047 & 0.058 & 0.55 & 1.27 & 0.32\\
2019-06-29 & 0.003 & 0.102 & 0.096 & 0.26 & 1.13 & 0.51\\
				\hline
				\multicolumn{7}{c}{N-band}\\
				 Date & \multicolumn{3}{c}{Correlated flux (\%)} & \multicolumn{3}{c}{Closure phase ($^\circ$)} \\
				 & $\sigma_\mathrm{r}$ & $\sigma_\mathrm{sys, s}$ & $\sigma_\mathrm{sys, l}$ & $\sigma_\mathrm{r}$ & $\sigma_\mathrm{sys, s}$ & $\sigma_\mathrm{sys, l}$  \\
				\hline
				2019-03-23 & 3.4 & 7.7 & 7.8 & 23.5 & 65.8 & 0.8\\
2019-05-06 & n.a. & n.a. & n.a. & 59.7 & 104.2 & 4.9\\
2019-06-26 & 5.0 & 16.1 & 21.7 & 39.7 & 78.0 & 19.8\\
2019-06-29 & 4.8 & 16.1 & 20.6 & 33.4 & 76.1 & 19.4\\
            \hline
			\end{tabular}
		\end{center}
	\end{table}

\begin{figure*}
   \centering
   \includegraphics[width=\hsize]{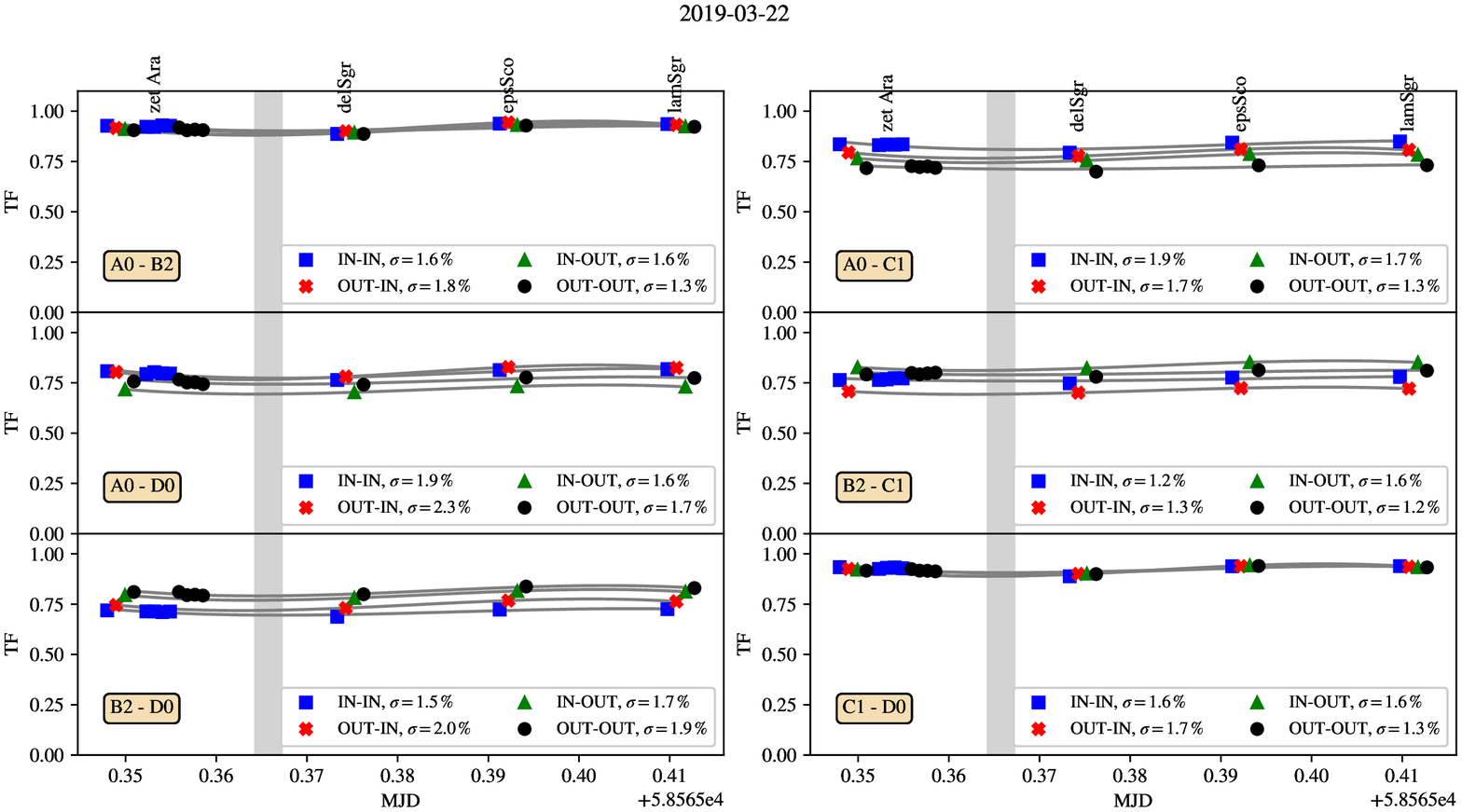}
      \caption{L-band transfer function during the night 2019-03-22. Grey shading indicates the observation time of HD 163296. The gray lines, one for each BCD configuration, are cubic polynomial fits to the points. In the captions, $\sigma$ is the relative standard deviation of the transfer function values.
              }
         \label{fig:TF_L_03_22}
  \end{figure*}
 
 \begin{figure*}
   \centering
   \includegraphics[width=\hsize]{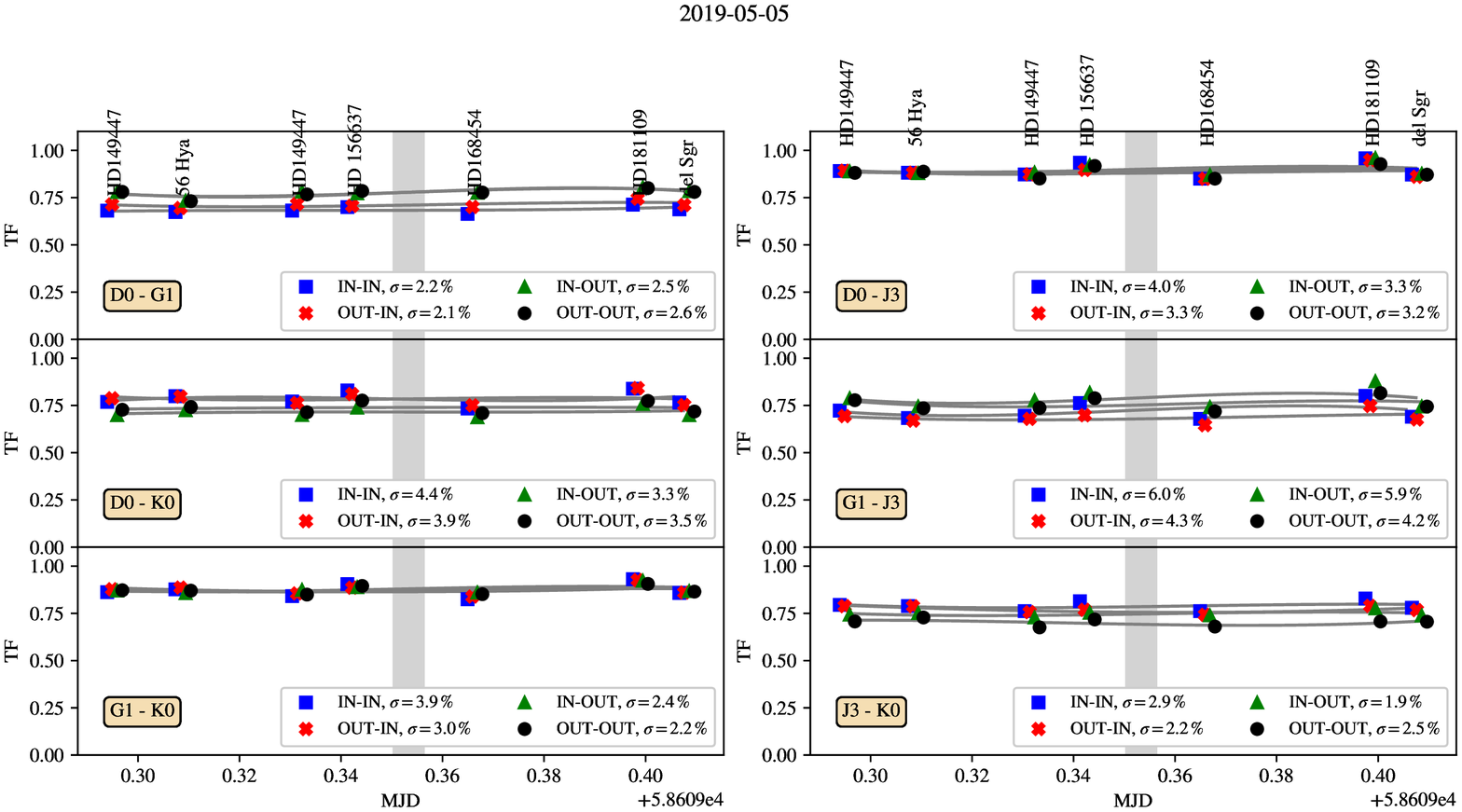}
      \caption{Same as Fig.~\ref{fig:TF_L_03_22}, but for the night 2019-05-05.
              }
         \label{fig:TF_L_05_05}
  \end{figure*}
 
 \begin{figure*}
   \centering
   \includegraphics[width=\hsize]{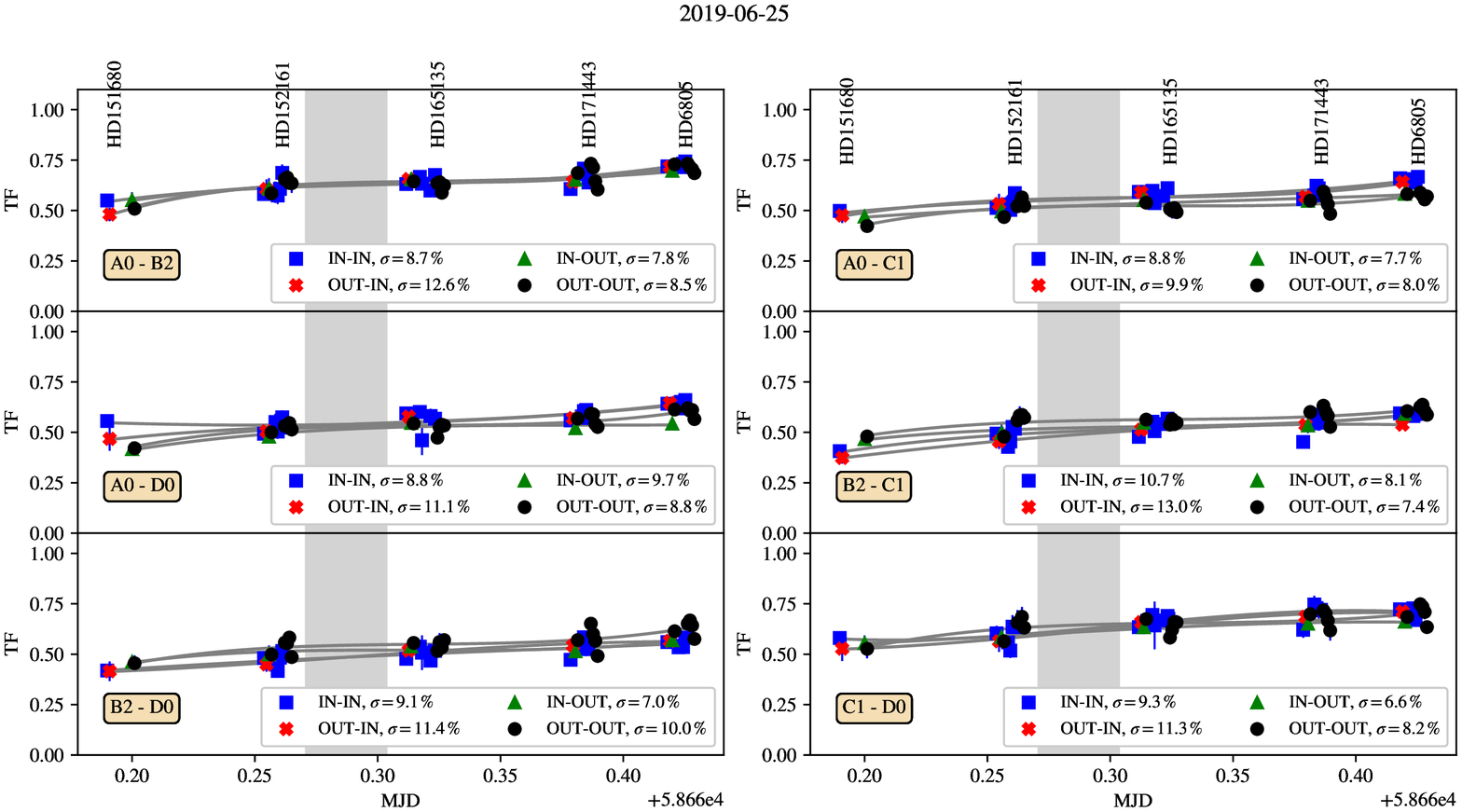}
      \caption{Same as Fig.~\ref{fig:TF_L_03_22}, but for the night 2019-06-25.
              }
         \label{fig:TF_L_06_25}
  \end{figure*}
  
  \begin{figure*}
   \centering
   \includegraphics[width=\hsize]{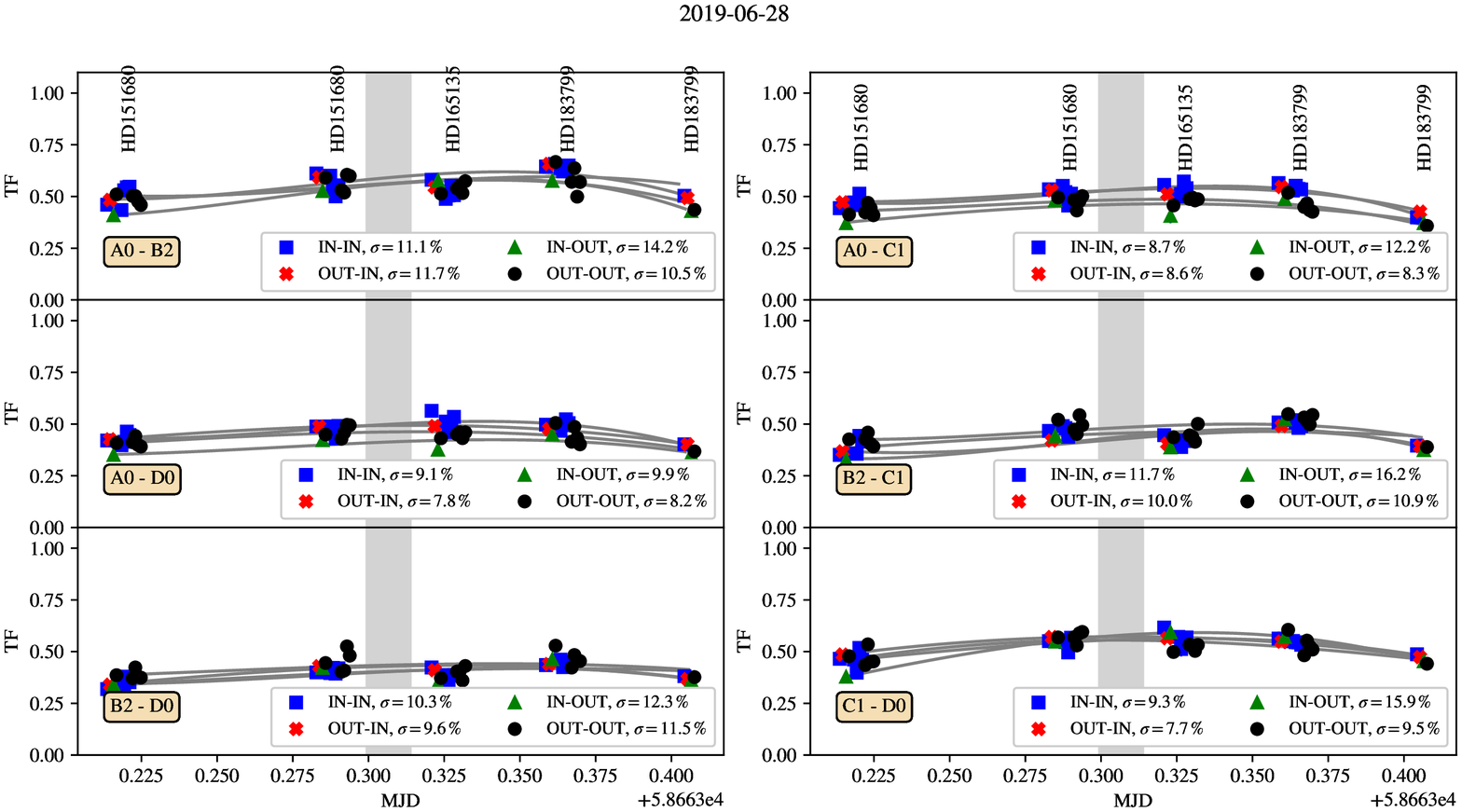}
      \caption{Same as Fig.~\ref{fig:TF_L_03_22}, but for the night 2019-06-28.
              }
         \label{fig:TF_L_06_28}
  \end{figure*}
  
The aim of this section is to assess the uncertainty of the calibrated data products. The MATISSE pipeline estimates errors by dividing a raw exposure into chunks, and reducing these chunks separately. Then, the error is taken as the standard deviation of the reduced data in the chunks. This estimation accounts for the instrument-related noise sources. However, a more significant source of error is our incomplete knowledge of the transfer function. During the typically $20-30$~min time lag between a science and calibrator observation the atmospheric conditions could change a lot, so the transfer function varies significantly. This error causes an uncertainty in the overall visibility or flux level of the spectrum, thus it is systematic in nature. In Figures~\ref{fig:TF_L_03_22}-\ref{fig:TF_L_06_28} we plot the L-band transfer function over time for each night, for each baseline, and for each BCD configuration. To assess the uncertainties in the calibrated data we employ the following: 
   \begin{itemize}
       \item We estimate the random noise ($\sigma_\mathrm{r}$) by subtracting the trend from the data, and taking the standard deviation of the residual signal. This error is the uncorrelated noise between the spectral channels.
       \item We estimate short-term systematic uncertainties ($\sigma_\mathrm{sys,s}$) by taking the standard deviation of the 4 or 8 exposures taken during the interferometric observation\footnote{The number of non-chopped exposures is $4 N$, where $N$ is the number of exposure cycles.}. The characteristic time scale corresponding to $\sigma_\mathrm{sys,s}$ is $4-8$~min.
       \item We estimate long-term systematic uncertainties ($\sigma_\mathrm{sys,l}$) by calibrating the science data with alternative calibrators. One calibrator is chosen before, the other after the science observation. Thus we have 3 calibrated data-sets: one with the original calibrator, and two with the alternative calibrators. Then we take the standard deviation of these data-sets. The characteristic time scale corresponding to $\sigma_\mathrm{sys,s}$ is $1-3$~hours.
   \end{itemize}
   
The average uncertainties (averaged over baselines and wavelengths) are listed in Table~\ref{tab:error} for each of the four observations. For N-band we show the relative errors on the correlated flux. For the L-band visibility and for the N-band correlated flux systematic uncertainties are larger than the random noise. Comparing the systematic errors, we see that $\sigma_\mathrm{sys,l} > \sigma_\mathrm{sys,s}$, indicating that hour-long variations of the atmosphere are larger than changes over a few minutes. The values for $\sigma_\mathrm{sys,l}$ in L-band are consistent with the scatter in the transfer function values, shown in the captions of the Figures~\ref{fig:TF_L_03_22}-\ref{fig:TF_L_06_28}. We note that the May data-set has the smallest absolute errors on visibility, however, in relative terms the March data-set is better. The origin of this difference is that the May data-set has significantly lower L-band visibilities (compared to the March data). In relative terms the March data has $\sigma_\mathrm{sys,l}$ values of $\approx2\%$, while the corresponding values for the May data are in the range of $2-6\%$. 

In the closure phases the largest uncertainty is the $\sigma_\mathrm{sys,s}$, which actually reflects instrumental effects (revealed by the beam commutation), not atmospheric variations. Averaging the exposures suppresses the instrumental noise, and results in greatly reduced systematic uncertainties. Furthermore, $\sigma_\mathrm{sys,l} < \sigma_\mathrm{sys,s}$, indicating that the variable atmosphere has a smaller impact on the closure phase uncertainty, than the instrumental effects. The most significant error source on the closure phase which remains after averaging the exposures is the random noise ($\sigma_\mathrm{r}$). It is still within $1^\circ$ in L-band, but quite large ($>20^\circ$) in N-band. As a summary for Table~\ref{tab:error}, the overall uncertainty in L-band visibility is $<0.02$ for the March and May data (taken in very good weather), and $>0.06$ for the June data which was recorded under unfavorable atmospheric conditions. For the N-band correlated flux the overall uncertainty is $<8\%$ for the March data, and $\sim$$20\%$ for the June data. 

The June data-sets are generally consistent with the March data, but having significantly larger uncertainties. As the baselines probed in March and June were very similar (on the small AT array), there is little added value of including the lower quality June data in the modeling. Thus, we do not use these data in our analysis at all. Additionally, the N-band data from May suffers from a bias affecting the correlated flux at flux levels close to the instrument sensitivity limit ($5-8$ Jy with ATs). The correlated fluxes of HD 163296 at $\gtrsim 50$~m baselines probed in May, based on earlier VLTI/MIDI observations, are expected to be less than $8$~Jy \citep{Varga2018}. Due to the bias we are unable to provide error estimates for the May N-band correlated fluxes, and we do not use these data in this study. 
   
When calculating our final averaged calibrated data we assign an uncertainty to each data point by combining the error  provided by the MATISSE pipeline and our $\sigma_\mathrm{sys,s}$ estimate. Thus, long-term systematics are not included in the error bars. Additionally, our error analysis does not encompass errors caused by the uncertainty in the calibrator diameter, and errors due to the uncertainty in the calibrator model spectrum. For our modeling we set conservative lower limits on the total uncertainties, which are $0.03$ for the L-band visibility, $1^\circ$ for the L-band closure phase, and $8\%$ for the N-band correlated flux. 
   
\section{Posterior distributions}

\begin{figure*}
   \centering
   \includegraphics[width=\hsize]{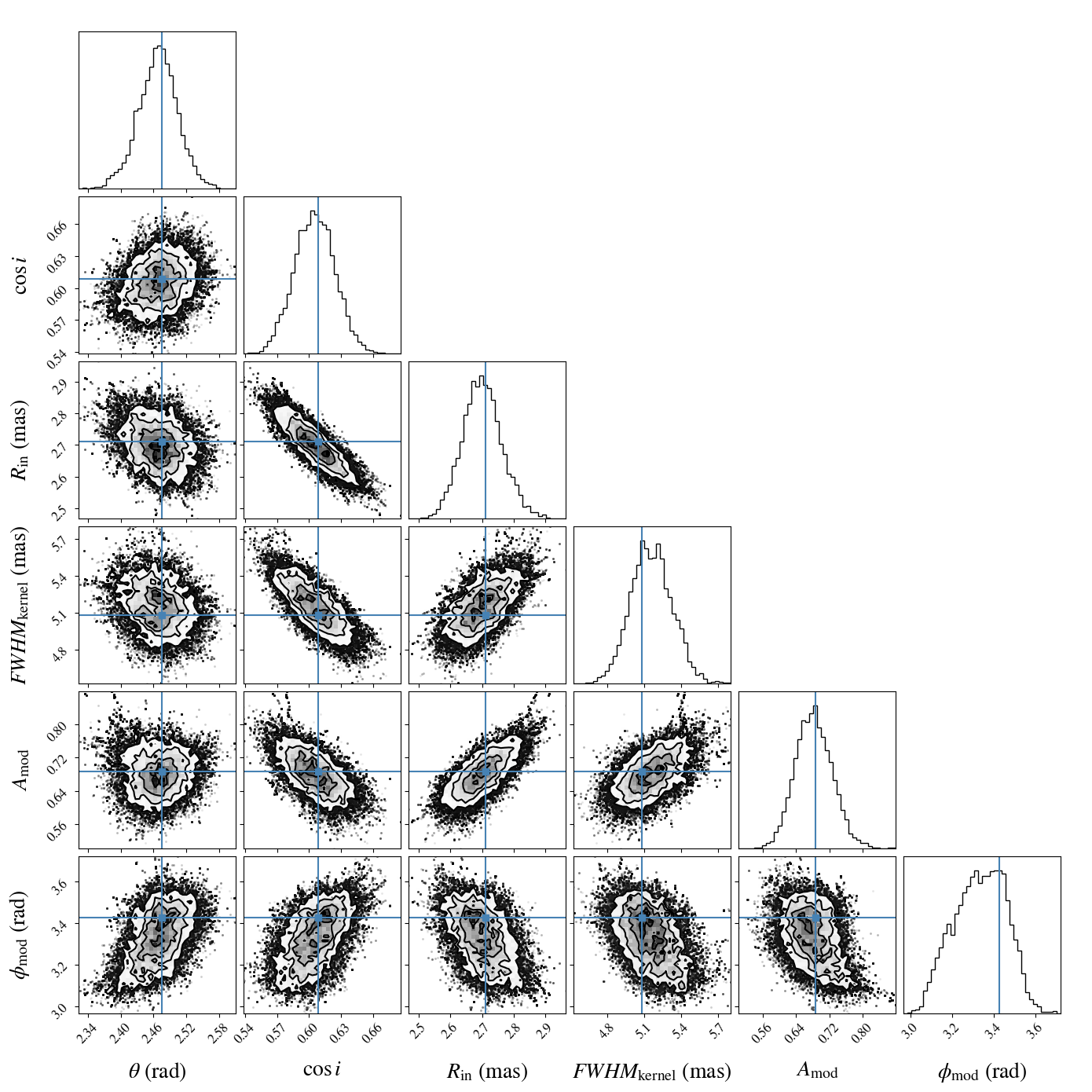}
      \caption{Posterior distributions for the smoothed ring model in L-band from our MCMC sampling. Blue lines represent the best-fit values.
              }
         \label{fig:cornerplot_L_ring}
  \end{figure*}

\begin{figure*}
   \centering
   \includegraphics[width=\hsize]{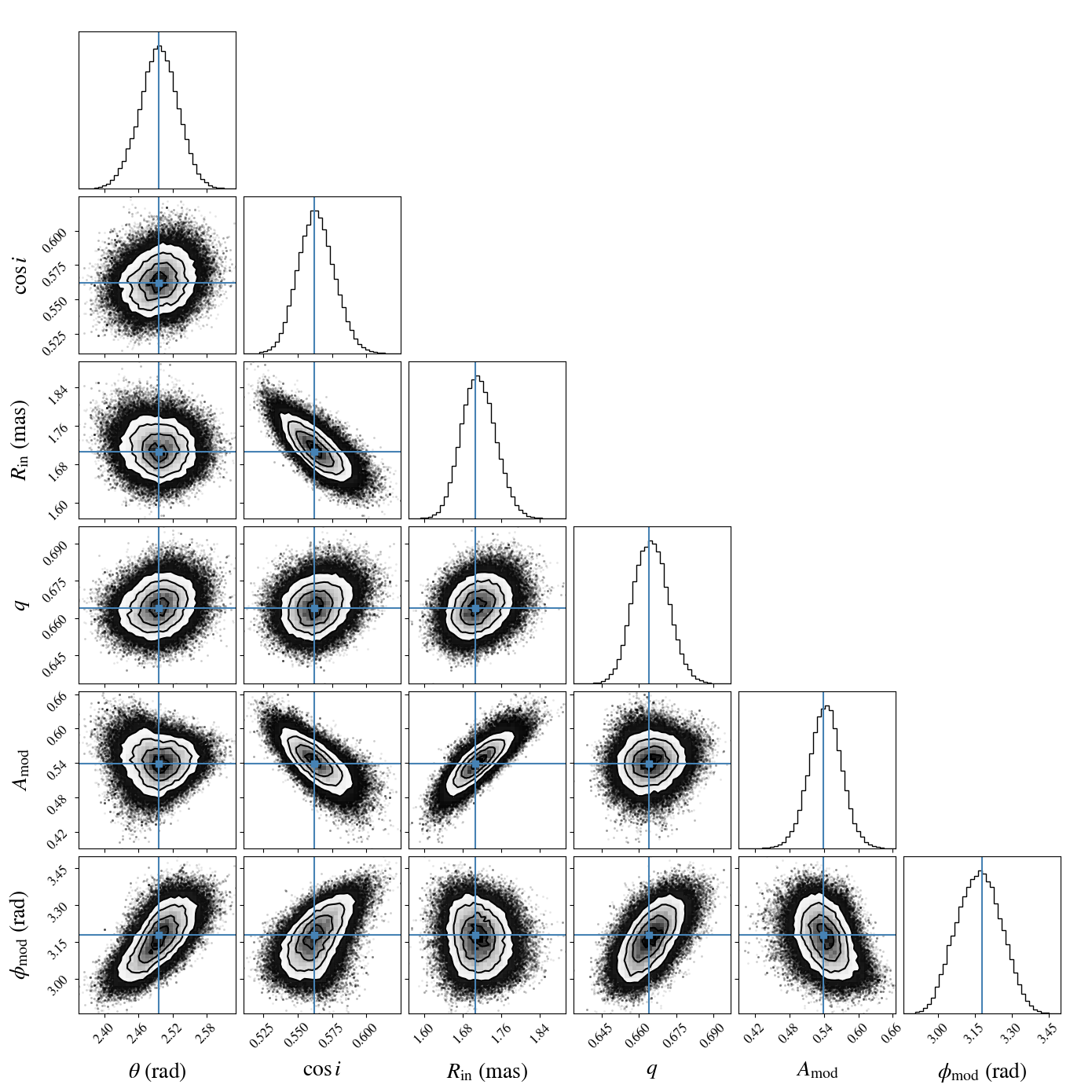}
   \caption{Same as Fig.~\ref{fig:cornerplot_L_ring}, but for the flat disk temperature gradient model.
              }
         \label{fig:cornerplot_L_flat_temp_grad}
  \end{figure*}

Figures \ref{fig:cornerplot_L_ring} and \ref{fig:cornerplot_L_flat_temp_grad} show the posterior distributions of the MCMC chain for the smoothed ring model and flat disk model, respectively. 

\section{Additional modeling}
\label{sec:gauss_model}

In order to explore whether a centrally symmetric model could fit the L-band data, we apply a model featuring a central 2D elliptical Gaussian. The central star, like in the previous L-band models, is represented as point source, with a fixed flux ratio. To account for asymmetry, we add a component which can be at an offset position. We first experimented with models where we used a Gaussian blob for the additional component. In these models the fitting converged towards a very small size for the Gaussian, without constraining a lower limit for the size. Thus, we choose to represent the the additional component as a point source. This model has 6 fitted parameters, just like our other models for the L-band. The central Gaussian is modeled with the following 3 parameters: the half width half maximum size ($HWHM_\mathrm{Gaussian}$), the axis ratio ($\cos i$), and the position angle of the major axis ($\theta$). The remaining 3 parameters are the coordinates of the additional point source ($x_\delta$, $y_\delta$), and its flux ratio with respect to the whole circumstellar emission ($f_\delta$). 

\begin{figure*}
   \centering
   \includegraphics[width=\hsize]{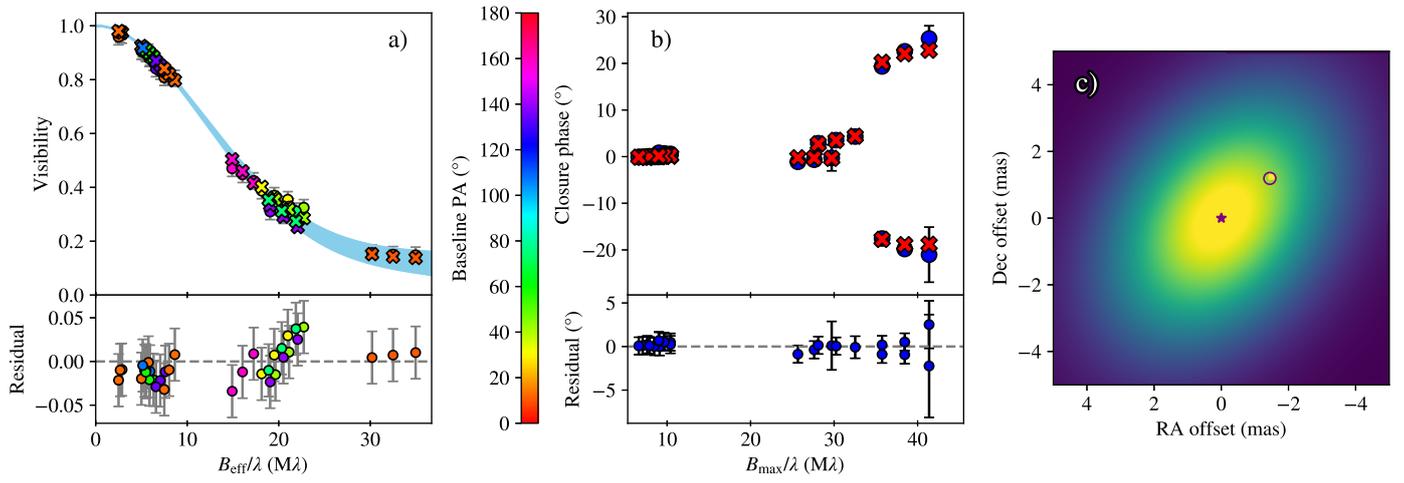}
   \caption{Same as Fig.~\ref{fig:model_fit_L_ring}, but with the Gaussian plus point source model. In panel c) we indicate the locations of the central star (star symbol) and of the additional point source (circle).
              }
         \label{fig:model_fit_L_gaussian_plus_delta}
  \end{figure*}

\begin{figure*}
   \centering
   \includegraphics[width=\hsize]{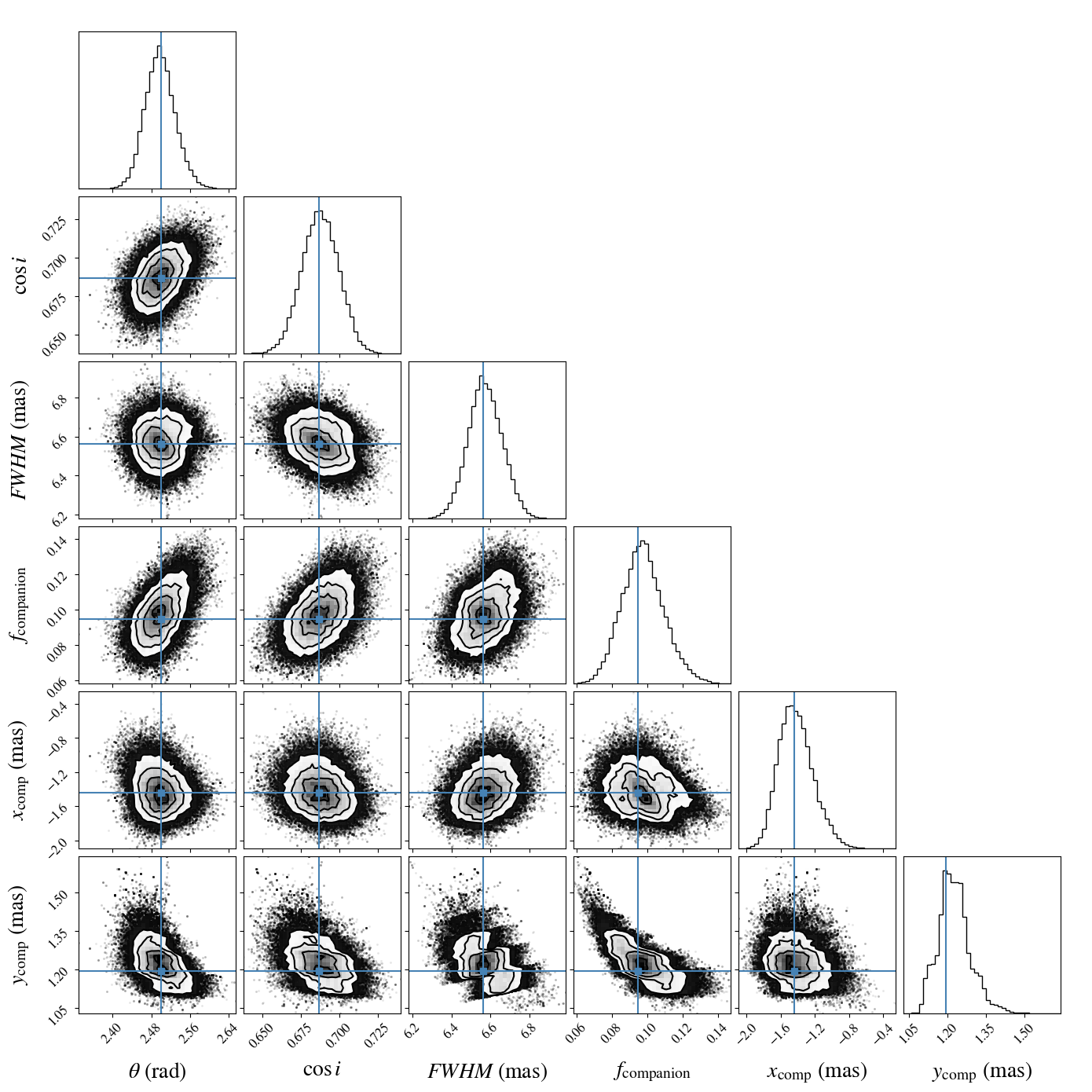}
   \caption{Same as Fig.~\ref{fig:cornerplot_L_ring}, but for the Gaussian plus additional point source model. 
              }
         \label{fig:cornerplot_L_gaussian_plus_delta}
  \end{figure*}

The fitting procedure was the same as described in Sect.~\ref{sec:modelfit}. The resulting fits are presented in Fig.~\ref{fig:model_fit_L_gaussian_plus_delta}, and the corresponding posterior distributions are shown in Fig.~\ref{fig:cornerplot_L_gaussian_plus_delta}. The fitted parameters are listed in Table~\ref{tab:res_L_Gauss}. 

\begin{table}
\caption{List of the best-fit parameters, half-light radii, and $\chi^2$-values in the Gaussian L-band modeling. }
\begin{center}
	\label{tab:res_L_Gauss}

\begin{tabular}{l c}
\hline \hline
 & Gaussian $+$ point \\
 & source model\\
\hline
$\theta$ ($\degr$)  &  $143.3^{+1.6}_{-1.9}$ \\
$\cos\,i$  &  $0.69^{+0.01}_{-0.01}$ \\
$HWHM_\mathrm{Gaussian}$ (mas)  & $3.28^{+0.05}_{-0.04}$ \\
$HWHM_\mathrm{Gaussian}$ (au)  & $0.332^{+0.005}_{-0.004}$ \\
$f_\delta$ & $0.09^{+0.01}_{-0.01}$ \\
$x_\delta$ ($\mathrm{mas}$)  & $-1.45^{+0.23}_{-0.18}$ \\
$y_\delta$ ($\mathrm{mas}$)  & $1.19^{+0.09}_{-0.02}$ \\
$x_\delta$ ($\mathrm{au}$)  & $-0.147^{+0.023}_{-0.018}$ \\
$y_\delta$ ($\mathrm{au}$)  & $0.120^{+0.009}_{-0.002}$ \\
\hline
$R_\mathrm{hl}$ (mas)  & $3.28$ \\
$R_\mathrm{hl}$ (au)  & $0.33$ \\
$\chi^2_V/N_V$ & $0.39$ \\
$\chi^2_\mathrm{CP}/N_\mathrm{CP}$ & $0.23$ \\
\hline
\end{tabular}

\end{center}
\end{table}

\end{appendix}

\end{document}